\documentclass[a4paper,fleqn,usenatbib,useAMS]{mnras}

\usepackage{amsmath}	% Advanced maths commands
\usepackage{amssymb}	% Extra maths symbols
\usepackage{multicol}        % Multi-column entries in tables
\usepackage{bm}		% Bold maths symbols, including upright Greek
\usepackage{pdflscape}	% Landscape pages
\usepackage[T1]{fontenc}
\usepackage{ae,aecompl}
\usepackage{txfonts}
\usepackage{longtable}
\usepackage{ulem}
\usepackage[pdftex]{graphicx}
\usepackage{hyperref}
\newcommand{\kms}{\,km\,s$^{-1}$} % kilometres per second
\def\H2O       {H$_2$O }

\def\h       {\ifmmode{^{\rm h}}\else{$^{\rm h}$}\fi}
\def\m       {\ifmmode{^{\rm m}}\else{$^{\rm m}$}\fi}
\def\s       {\ifmmode{^{\rm s}}\else{$^{\rm s}$}\fi}
\def\deg     {\ifmmode{^{\circ}}\else{$^{\circ}$}\fi}

\def\decdeg  {\ifmmode{{\rlap.}^{\circ}} \else ${\rlap.}^{\circ}$\fi}
\def\decs    {\ifmmode{{\rlap.}^{\rm s}} \else ${\rlap.}^{\rm s}$\fi}
\def\decas   {\ifmmode{{\rlap.}{''}}\else{${\rlap.}{''}$}\fi}

\def\arcsec  {$^{\prime}$}
\def\arcsec  {$^{\prime\prime}$}

\def\Ta        {$T^\ast_A$}

\title{\textit{Water Maser Survey towards off-plane O-rich AGBs around the orbital plane of the Sagittarius Stellar Stream}}
\author[Yuanwei Wu]{Yuanwei Wu$^{1}$\thanks{\href{yuanwei.wu@ntsc.ac.cn}{yuanwei.wu@ntsc.ac.cn}}, Bo Zhang$^{2}$, Jingjing Li$^{3}$, Xing-Wu Zheng$^{4}$
\\
$^{1}$National Time Service Center, Chinese Academy of Sciences, Xi'an 710600, China\\
$^{2}$Shanghai Astronomical Observatory, Chinese Academy of Sciences, Shanghai 200030, China\\
$^{3}$Purple Mountain Observatory, Chinese Academy of Sciences, Nanjing 210033,China\\
$^{4}$Department of Astronomy, Nanjing Univiersity 210093, China}
\date{Last updated 2022 January 25}
\pubyear{2022}
\begin{document}
\label{firstpage}
\pagerange{\pageref{firstpage}--\pageref{lastpage}}
\maketitle

\begin{abstract}
A 22 GHz water maser survey was conducted towards 178 O-rich AGB stars with the
aim of identifying maser emission associated with the Sagittarius stellar
stream. In this survey, maser emissions were detected in 21 targets, of which
20 were new detections. We studied the Galactic distributions of H$_2$O and SiO
maser-traced AGBs towards the Sgr orbital plane, and found an elongated
structure towards the ($l$, $b$)~$\sim$~(340$^{\circ}$, 40$^{\circ}$)
direction. In order to verify its association with the Sagittarius tidal
stream, we further studied the 3D motions of these sources, but found,
kinematically, these maser-traced AGBs are still Galactic disc sources rather
than Stream debris. In addition, we found a remarkable outward motion,
$\sim$50~ km~s$^{-1}$ away from the Galactic center of these maser-traced AGBs,
but with no systermatic lag of rotational speed which were reported in 2000 for
solar neighborhood Miras.

\end{abstract}


%\begingroup
%\let\clearpage\relax
%\tableofcontents
%\endgroup

\section{Introduction} \label{sec:intro}

It is well known that our Milky Way is a barred spiral galaxy. With in the past
decade, the spiral structure and kinematics of the Milky Way disc have been
well studied by measuring proper motions and parallaxes of interstellar masers
\citep{2019ApJ...885..131R}. Astronomical masers are bright and compact radio
sources, thus are very good targets for high-accuracy VLBI astrometry
\citep{2014ARA+A..52..339R}. Actually, they are so bright that they even have
been detected in external galaxies. Apart from spiral features in the disc,
there are also founded the large scale and tail-like features, the so called
stellar streams, in the halo region of the Milky Way. 

Stellar streams can be produced by accretion and merges of our Galaxy with
satellite dwarf galaxies, with Galactocentric distances ranging from $\sim$ 10
kpc to more than 100 kpc.  The most prominent and well studied stream is the
Sagittarius tidal stream (hereafter Sgr stream), which was produced by the
interaction of the Milky Way with its nearest satellite, the Sagittarius Dwarf
Spheroidal Galaxy (Sgr dSph).  The existence of the Sgr stream was firstly
anticipated by \citet{1995MNRAS.275..429L} by investigating the kinematics of
global clusters.  \citet{2001ApJ...551..294I} firstly identified this structure
from carbon stars. After then, a variety of tracers were used to characterize
the stream, including M giants \citep{2003ApJ...599.1082M}, RR Lyrae
\citep{2006AJ....132..714V, 2013ApJ...765..154D}, horizontal branch stars
\citep{2011ApJ...731..119R, 2012ApJ...751..130S}, red clump stars
\citep{2010ApJ...721..329C, 2012AJ....144...18C}, Carbon stars
\citep{2001ApJ...551..294I, 2015MNRAS.453.2653H}, upper main-sequence and
main-sequence turn-off stars \citep{2006ApJ...642L.137B, 2012ApJ...750...80K,
2013ApJ...762....6S}. In general, most observational evidences
of the Sgr stream are from the northern hemispheric leading (L1) arm and the
southern hemispheric Trailing (T1) arm, as these are the most young and
remarkable features of the Sgr stream, with distance ranging from 15 to 100~kpc
\citep{2020A+A...638A.104R}.

Theorists believe that many of global morphological features seen in the
Galactic disc, for example Monoceros ring \citep{2002ApJ...569..245N,
2008ApJ...673..864J}, can be explained as perturbations by Sagittarius dwarf
galaxy \citep{2013MNRAS.429..159G}.  In our previous studies on the radical
velocities of thick disk SiO masers, we identified the large-scale peculiar
motions in the Perseus arm, the group C, 0.3$<Z<$0.8~kpc, $R>9$~kpc region,
\citep{2018MNRAS.473.3325W}, which are also seen in the LAMOST and Gaia red
clump sample as the asymmetrical kinematics of mono-ageo polulations
\citep{2020MNRAS.491.2104W}. Meanwhile, within 2~kpc of the solar neigborhood,
\cite{2020NatAs...4..965R} found three conspicuous and narrow episodes of
enhanced star formation occurred 5.7, 1.9 and 1.0 Gyr ago.  Such repeated star
forming history  is explained as the periodical perturbations due to
interaction between the Sagittarius dwarf galaxy and the Milky Way disc.

The orbital period of the Sgr dwarf galaxy is around 1 Gyr, 3D
simulations indicates a miminum age of 2 or even 3 orbital period
\citep{2010ApJ...714..229L, 2010MNRAS.408L..26P}, therefore, there should
exists the L2, T2, or even L3, T3 features in the halo or even solar
neigborbhood \citep{2010ApJ...714..229L, 2021arXiv211202105R}. The 3D
simulations suggest the L2 feature could be close to the solar neighborhood, or
even 'direct hit' on the solar neighborhood \citep{2004AJ....128..245M}.
Taking the Monoceros ring, the solar neigborhood star formation history, and
the 3D simulations into account, yields the possibility that the Sgr dwarf
galaxy may cross the Milky Way plane not far away from the solar neighborhood.

%\citet{2019A+A...626A.112M} compiled a halo AGB M star catalog, which indicates
%a strong concentration of M-type AGBs along the apparent orbital plan of the
%Sgr tidal arm.  Especially, \citet{2019A+A...626A.112M} found about three times
%more M-type stars than C-type stars. suggest M-type AGB stars which maybe older
%than C stars can be a good tracer to the older components Sgr stream.

%All stars between 1-8 $M_\odot$ will pass through the asymptotic giant
%branch (AGB) phase, at their late evolution stages. The AGB stars have C-O
%cores surrounded by a He-burning shell and a H-burning shell, and pulsating
%ircumstellar envelopes. In the circumstellar evvelopes of O-rich (C/O$<1$)
%AGBs, exists circumstellar SiO and H$_2$O maser emissions which are bright and
%compact for radio interferometry .

\citet{2018MNRAS.473.3325W} conducted SiO masers survey towards a WISE selected
O-rich AGB stars towards the Sgr orbital plane, in the SiO maser survey, 45 SiO
maser are detected. Given water masers are generally brighter than SiO masers,
it could be more possible to detect H$_2$O masers at larger distances, thus
after SiO maser survey, we further conducted 22 GHz water maser survey on the
similar WISE selected O-rich AGB stars towards the Sgr orbital plane. Here we
report the result of the 22 GHz water maser survey. In Section \ref{sec:obs} we
present sample and observations of this survey. In Section \ref{sec:results} we
present the result of this survey. Discussions on the Galactic locations and
kinematics of both water and previous reported SiO masers are presented in
Section \ref{sec:discuss}. A summary is given in Section \ref{sec:summary}.

\section{Sample and Observation} \label{sec:obs}
\subsection{Sample}  \label{subsec:sample}

\begin{figure*}[H]
\includegraphics[width=16cm]{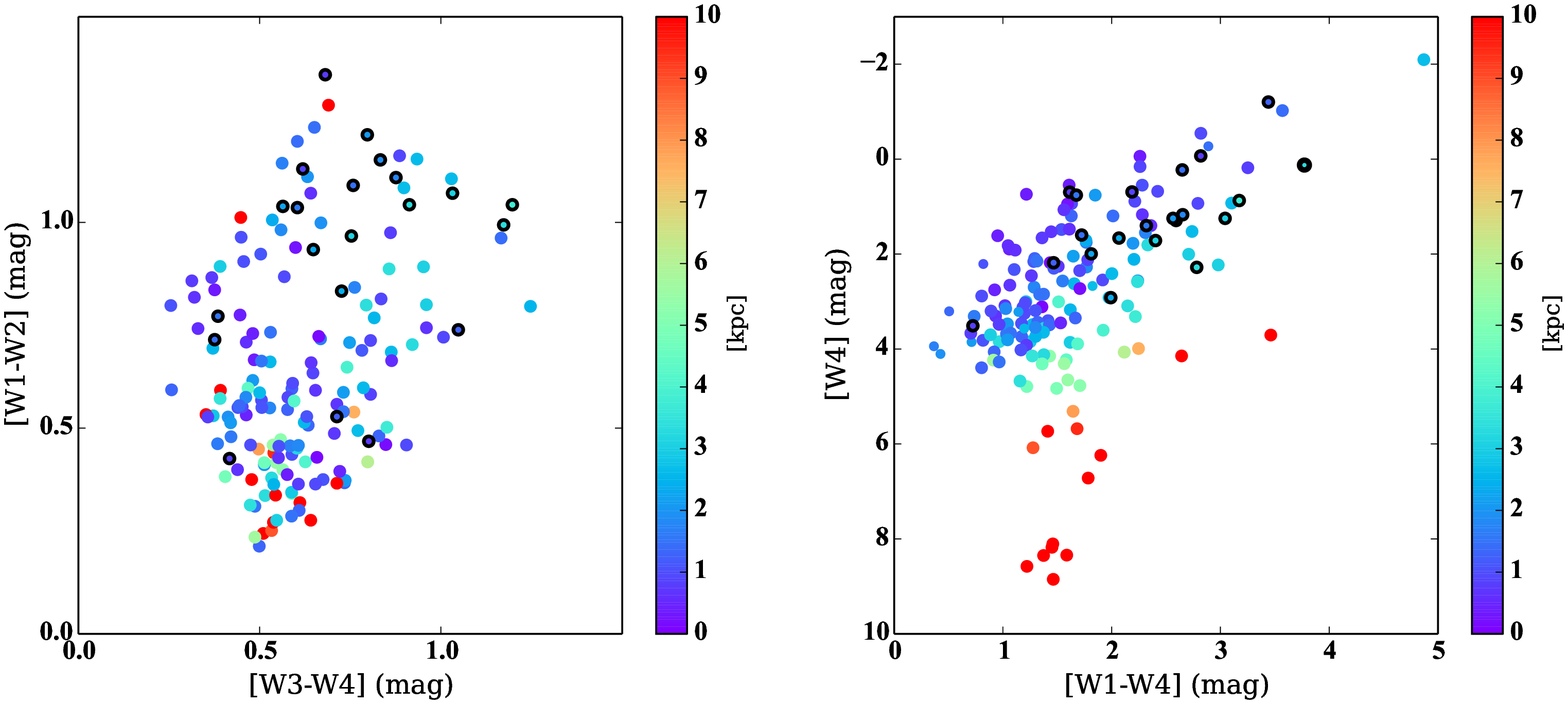}
 \caption{\textit{Left panel:} WISE colour-colour diagram ([$W3$-$W4$] versus
[$W1$-$W2$]).  \textit{Right panel}:
WISE colour-magnitude diagram([$W1$-$W4$] versus $W4$) diagram Colour of
symbols denote distances, Symbols with black edges are sources with detection
of 22 GHz water masers. \label{fig-1}} \end{figure*}

The O-rich AGBs are selected from the Wide-field Infrared Survey Explorer
(WISE) all-sky catalogue by color-color diagram following the same criteria of
the 2018 SiO maser survey \citep{2018MNRAS.473.3325W}. In Figure \ref{fig-1}
shown are the WISE [W3-W4] versus [W1-W2] color-color diagram, and the [W1-W4]
versus [W4] color-magnitude diagram, with symbol colors denote the distance of
source. In total, we searched for 22 GHz water masers toward 176 targets,
including 67 Miras, 49 Mira candidates, and 60 other types of long period
variables (LPV) or LPV candidates. 

\begin{figure*}[H]
\includegraphics[width=16cm]{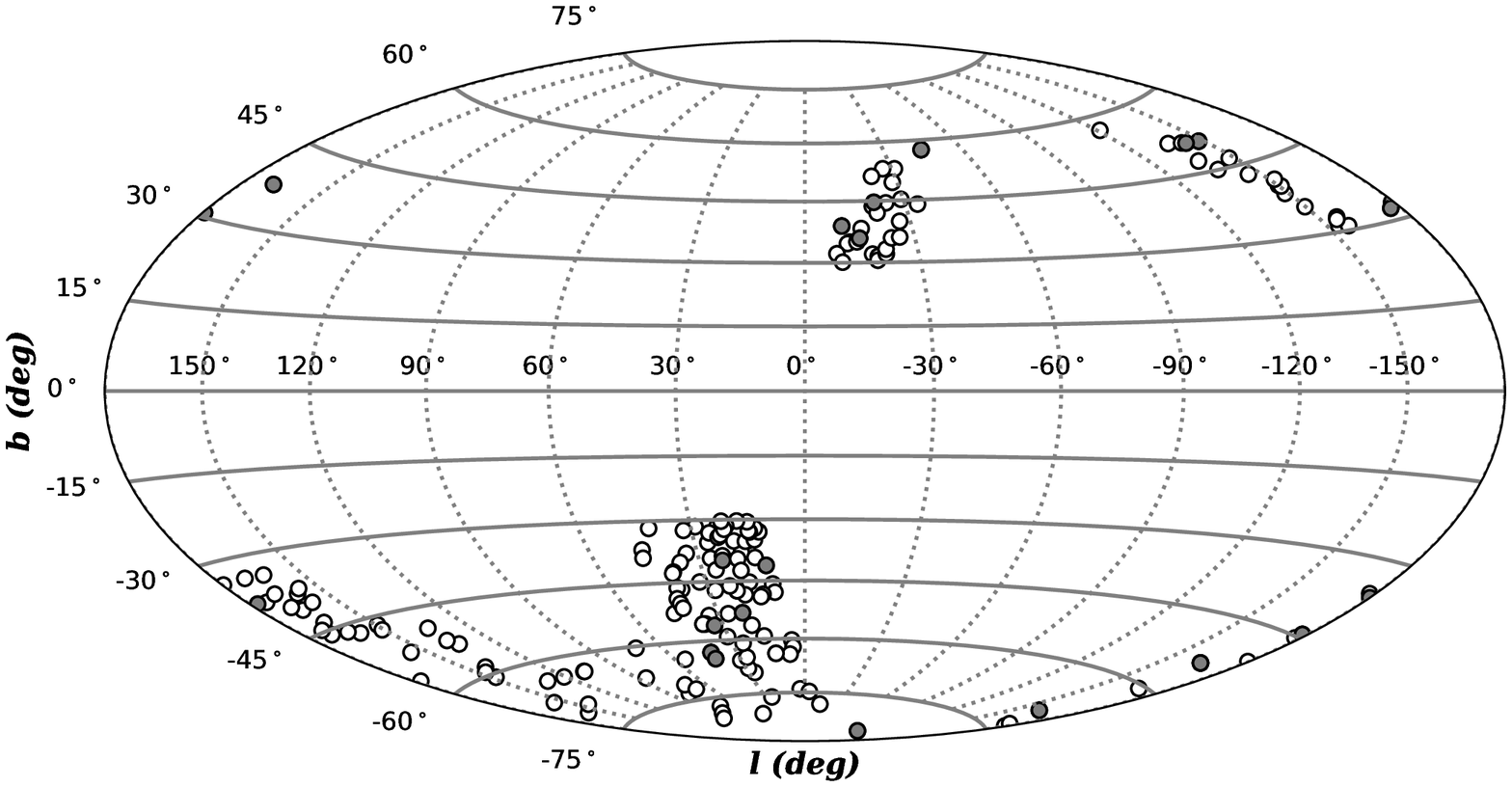}
\includegraphics[width=16cm]{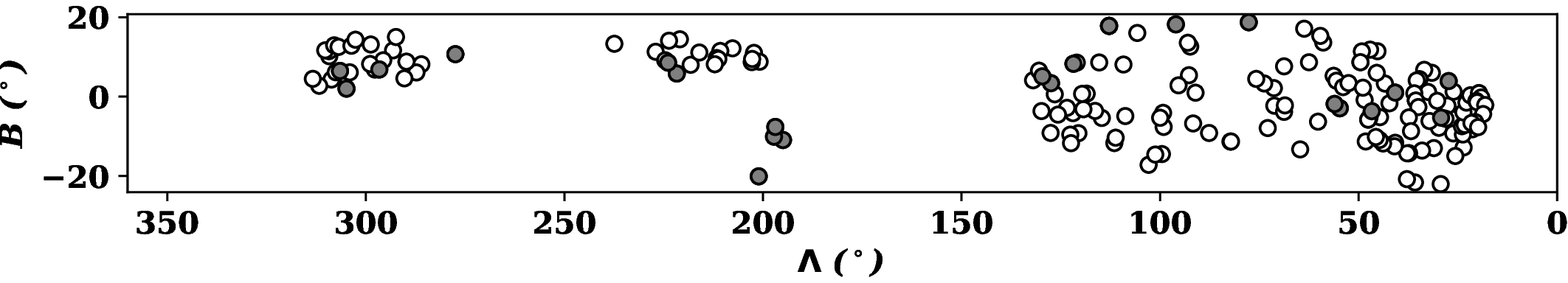}
\caption{\textit{Top panel}: Galactic distribution of observed targets, filled grey circles
indicates sources with detection of SiO masers. 
\textit{Bottom panel}: Same as top panel but in the Stellar Stream coordinate
system.  
\label{fig-2}} 
\end{figure*}

In Figure \ref{fig-2}, we present the locations of our target sources in both
the Galactic coordinates and Sgr stellar stream coordinates, where filled grey
circles indicates sources with detection of maser emissions. Source details,
i.e., source name (Galactic coordinate notation), WISE name (equatorial
coordinate notation), Bayer designation names, star type (queried from the
SIMBAD database), period (queried from GCVS and AAVSO variable star catalog),
and detection results are all listed in Table \ref{tab-A1}.

\subsection{Observation}\label{subsec:obs} 

We carried out observations of the H$_2$O 6$_{16} - 5_{23}$ (22.235080~GHz)
maser lines with the Nobeyama 45-m telescope\footnote{The 45-m radio telescope
is operated by Nobeyama Radio Observatory, a branch of National Astronomical
Observatory of Japan.} in May 2016, and the Tidbinbilla DSS-43 70m
telescope\footnote{The 70-m radio telescope are operated by the Canberra Deep
Space Communication Complex, part of NASA's Deep Space Network.} in November
2016 and March 2017. For the 176 observed O-rich AGB stars, 49 were observed
with Nobeyama 45m telescope and the other 127 sources were observed with the
Tidbinbilla DSS-43 70m telescope. It is noted that, due to the limitation of
Nobeyama observation time, and the much higher maser detection rate of Mira
type stars, for the 49 sources observed by Nobeyama 45m, 47 are Mira type
stars.     

During Nobemaya 45-m observation sessions, pointings were checked every 2 hours
using known sources of strong H$_2$O maser emission. The half power beam width
of Nobeyama 45m was 75\arcsec\ at 22 GHz, with an aperture efficiency of 0.56.
The backend was set with bandwidth of 63 MHz and channel spacing of 15.26 kHz,
covering a velocity range of $\pm$420~\kms\ with a velocity spacing of 0.20
\kms.  System temperatures  were within 100 to 190 K. The integration time per
source was 10-30 minutes giving a 1$\sigma$ level of 0.04-0.08K. The data were
calibrated using the chopper wheel method, which corrected for atmospheric
attenuation and antenna gain variations to yield an antenna temperature \Ta.
The conversion factor from the antenna temperature, \Ta\, in units of K, to
flux density in units of Jy, was about 2.73Jy~K$^{-1}$.

During Tidbinbilla 70-m observation sessions, pointings were checked every 2
hours using strong H$_2$O maser source Orion KL. The half power beam width of
Tidbinbilla 70-m was  48\arcsec\ at 22 GHz. The 17-27 GHz dual horn receiver is
used to record the dual polarized maser emissions \citep{2019JAI.....850014K}.
The SAO Spectrometer was set with a bandwidth of 1~GHz at a spetral resolution
of 31.25 kHz or 0.42 \kms. The system temperature during observations were
within 60 to 100 K. The integration time per source was 3 minutes give a
1$\sigma$ level of 0.03-0.06K. The data were calibrated using the chopper wheel
method, which corrected for atmospheric attenuation and antenna gain variations
to yield an antenna temperature \Ta.  The conversion factor from the antenna
temperature, \Ta\, in units of K, to flux density in units of Jy, was about
1.5Jy~K$^{-1}$.

\section{Survey Results} \label{sec:results}

\begin{table*}
%\footnotesize
\setlength\tabcolsep{3pt}
\caption{List of detections of H$_2$O maser.
\label{tab-1}}
\begin{tabular}{rrlccrccccrc}
\hline
     Source & Other & Star   & R.A.(J2000) & DEC.(J2000) & $V_{\rm LSR}\quad$ &  T$^\ast_A$(peak) &  $\Delta v$            & Int. Flux       & rms & Telescope & SiO  \\
     Name   & Name  & Type   & (h:m:s)     & (d:m:s)     & (km s$^{-1}$)  & (K)                  &(km s$^{-1}$) & (K km s$^{-1}$)           & (K) &           & det \\
\hline
      G011.159$-$41.196  & X Mic     & Mi*  & 21 04 36.85 & $-$33 16 47.3 &    16.2  & 8.62  & 0.73  &  7.16  &  0.19 &NRO45m &  Y \\
      G021.513$-$53.023  & S PsA     & Mi*  & 22 03 45.83 & $-$28 03 04.2 & $-$98.0  & 1.36  & 0.90  &  1.69  &  0.10 &NRO45m &  Y \\
      G023.376$-$39.816  & V Cap     & Mi*  & 21 07 36.63 & $-$23 55 13.4 & $-$23.4  & 5.17  & 1.94  &  9.74  &  0.17 &NRO45m &  Y \\
      G033.245$-$56.048  & RT Aqr    & Mi*  & 22 23 12.94 & $-$22 03 25.5 & $-$29.1  & 2.16  & 1.08  &  3.50  &  0.10 &NRO45m &  N \\
      G041.307$-$63.037  & S Aqr     & Mi*  & 22 57 06.46 & $-$20 20 35.7 & $-$52.8  & 2.90  & 1.51  &  4.85  &  0.11 &NRO45m &  N \\
      G041.346$-$64.747  & MN Aqr    & Mi*  & 23 04 00.56 & $-$20 54 24.0 & $-$22.4  & 2.21  & 1.08  &  2.81  &  0.12 &NRO45m &  N \\
      G168.980$+$37.738  & X UMa     & Mi*  & 08 40 49.50 & $+$50 08 11.9 & $-$83.9  & 1.37  & 0.78  &  1.13  &  0.07 &NRO45m &  Y \\
      G177.272$-$37.906  &           & Mi*? & 03 32 32.90 & $+$07 25 32.3 & $-$40.6  & 0.30  & 0.72  &  0.57  &  0.04 &TID70m &  $\dagger$ \\
      G179.379$+$30.743  &           & Mi*  & 08 05 03.70 & $+$40 59 08.1 &  $-$6.9  & 0.14  & 7.72  &  0.75  &  0.04 &NRO45m &  N \\
      G181.889$-$44.366  &           & Mi*  & 03 22 31.61 & $+$00 31 48.0 &    32.3  & 1.50  & 1.97  &  5.57  &  0.04 &TID70m &  $\dagger$ \\
      G180.069$-$36.185  & V1083 Tau & Mi*  & 03 43 43.90 & $+$06 55 30.5 &    60.5  & 0.95  & 1.84  &  3.96  &  0.04 &TID70m &  Y \\
      G180.829$+$32.784  & W Lyn     & Mi*  & 08 16 46.88 & $+$40 07 53.3 & $-$24.4  & 0.20  & 1.60  &  0.40  &  0.04 &NRO45m &  Y \\
      G195.025$-$53.735  & SS Eri    & Mi*  & 03 11 53.14 & $-$11 52 32.4 &    32.4  & 7.53  & 0.75  &  6.30  &  0.10 &NRO45m &  Y \\
      G198.593$-$69.596  & RY Cet    & Mi*  & 02 16 00.08 & $-$20 31 10.5 &     0.4  & 1.42  & 1.07  &  1.80  &  0.11 &NRO45m &  Y \\
      G211.919$+$50.661  & V Leo     & Mi*  & 10 00 01.99 & $+$21 15 43.9 & $-$26.6  & 1.69  & 2.17  &  3.48  &  0.07 &NRO45m &  Y \\
      G217.372$+$50.948  &           & LPV? & 10 05 58.80 & $+$18 06 04.9 & $-$40.3  & 0.98  & 0.78  &  2.02  &  0.06 &TID70m &  $\dagger$ \\
      G248.071$-$84.665  & U Scl     & Mi*  & 01 11 36.38 & $-$30 06 29.4 & $-$10.8  & 2.27  & 1.82  &  4.79  &  0.14 &NRO45m &  Y \\
      G315.565$+$57.522  & VY Vir    & Mi*  & 13 18 30.52 & $-$04 41 03.2 &    70.9  & 0.31  & 1.25  &  0.34  &  0.06 &NRO45m &  Y \\ 
      G339.224$+$44.663  & KS Lib    & Mi*  & 14 32 59.87 & $-$10 56 03.2 &    67.3  & 0.35  & 1.80  &  1.38  &  0.03 &TID70m &  Y \\
      G345.104$+$35.879  &           & Mi*  & 15 08 54.49 & $-$15 29 51.0 & $-$59.5  & 0.72  & 0.83  &  1.88  &  0.06 &TID70m &  N \\
      G349.658$+$38.897  & TT Lib    & Mi*? & 15 12 23.62 & $-$10 51 51.7 & $-$58.1  & 0.45  & 0.61  &  0.98  &  0.04 &TID70m &  $\dagger$ \\
\hline
\multicolumn{12}{l}{{{\bf Note:} Column 1 are ID of sources; Column 2 are Galactic coordinate notated source names; column 3 and 4 are Bayer designation names of}}\\
\multicolumn{12}{l}{{variables and stellar types; column 5 and 6 are equatorial coordinates; column 7 and 8 are $V_{\rm LSR}$ and distance; column 9, 10, 11, 12 are}}\\
\multicolumn{12}{l}{{peak antennas temperatures (in unit of K), line width (in unit of km s$^{-1}$), integrated flux density (in unit of K km~s$^{-1}$), and 1$\sigma$}}\\
\multicolumn{12}{l}{{rms of spectra; column 13 are telescopes used for observations. In column 14, Y and N denote detection and non-detection of SiO maser,}}\\
\multicolumn{12}{l}{{$\dagger$ denote sources that are still not covered by any SiO maser survey.}}\\
\end{tabular}
\end{table*}

In Table \ref{tab-1}, we list the H$_2$O masers detected in this survey. In
total, we detected water maser emissions from 21 objects. We cross checked
these sources with the database\footnote{https://maserdb.net} of astrophysical
masers \citep{2019RAA....19...34S}, except for G211.919+50.661, which has been
reported the detection of 22 GHz maser emissions \citep{2001CRLRv..47..107T},
all the other 20 H$_2$O masers emissions were detected for the first time. We
further checked the stellar type of these 21 sources with the SIMBAD
astronomical database\footnote{https://simbad.u-strasbg.fr/simbad/}, for
stellar types of these 21 sources, apart from one source (G179.379+30.743)
without known stellar type, one source (G217.372+50.948) is classified as long
period variable candidate (LPV?), and 2 Mira candidates (G177.272-37.906 and
G349.658+38.897), all others are confirmed as Miras. In summary, the detection
rate of this water maser survey was 11.9\% for the whole sample, and 25.8\% for
the Mira type subsample.

\begin{figure*}
\includegraphics[width=5.0cm]{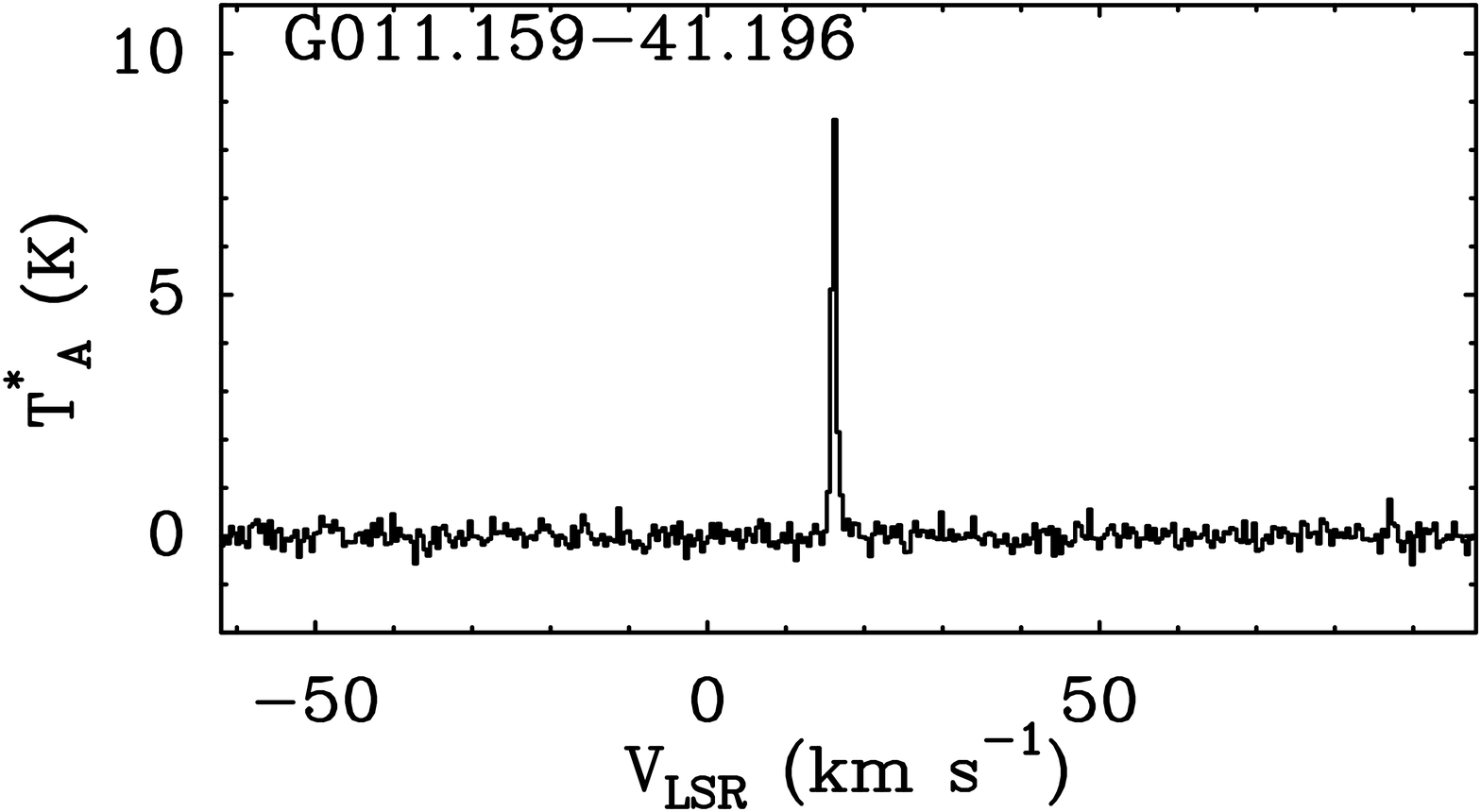}
\includegraphics[width=5.0cm]{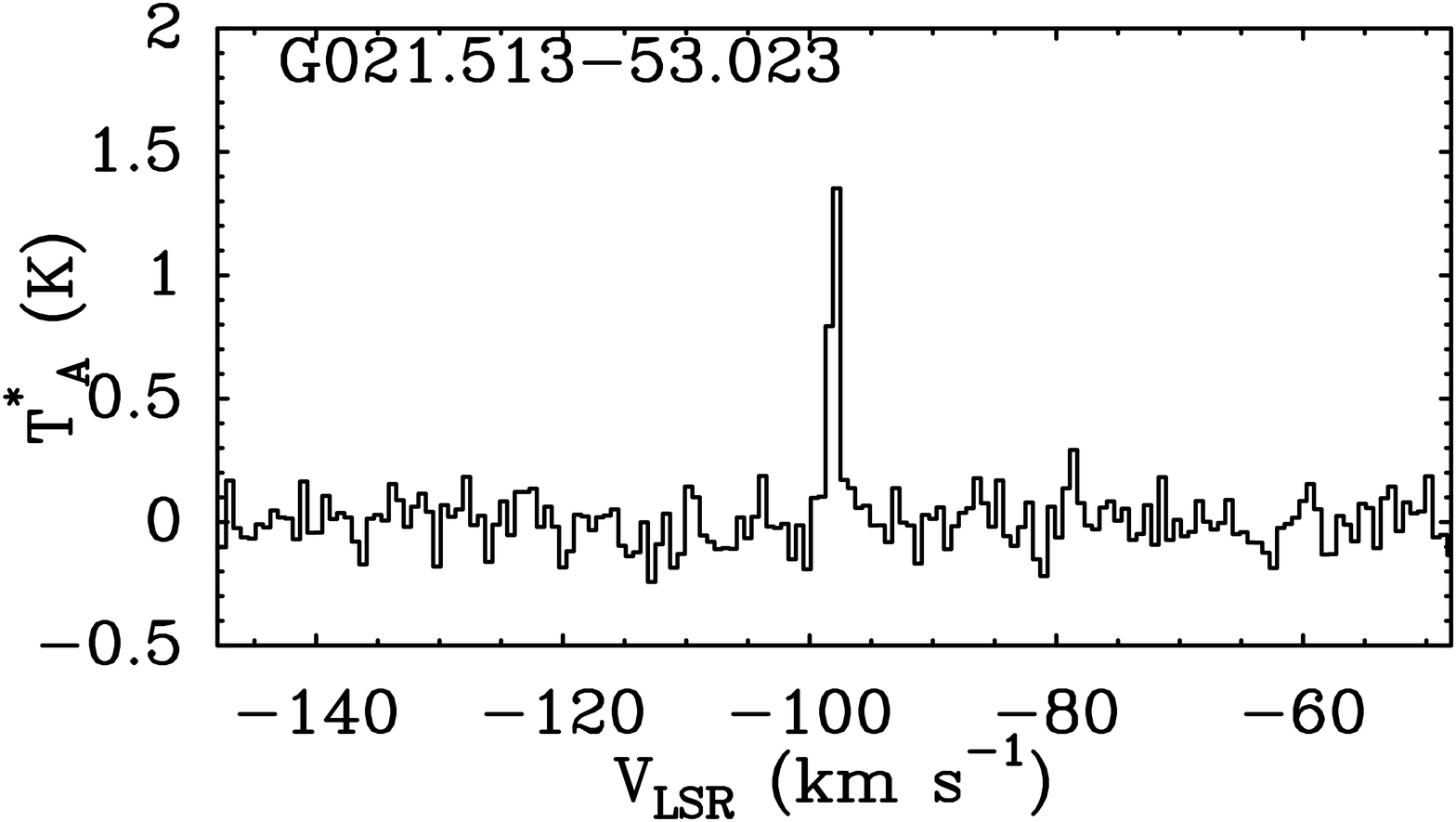}
\includegraphics[width=5.0cm]{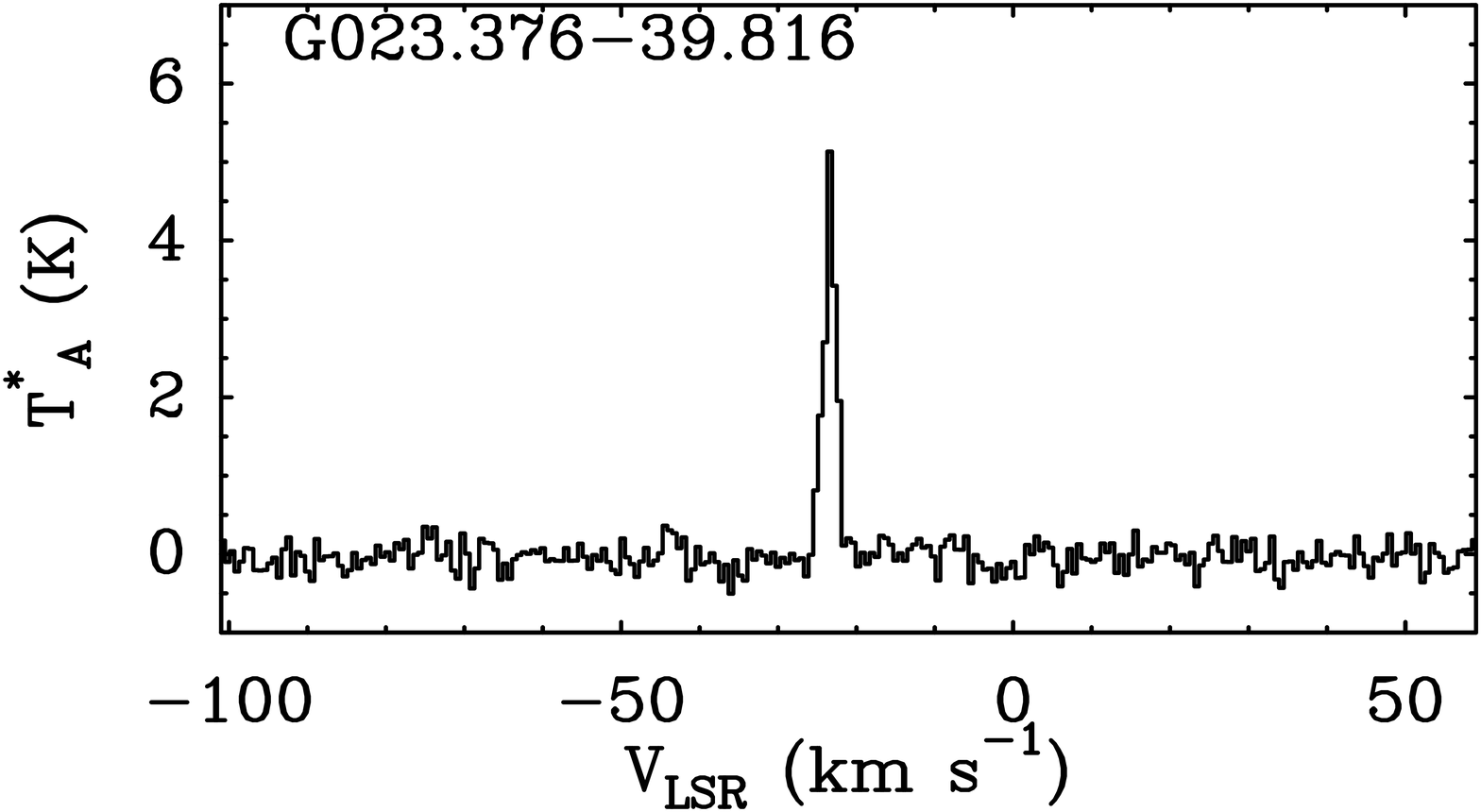} \\
\includegraphics[width=5.0cm]{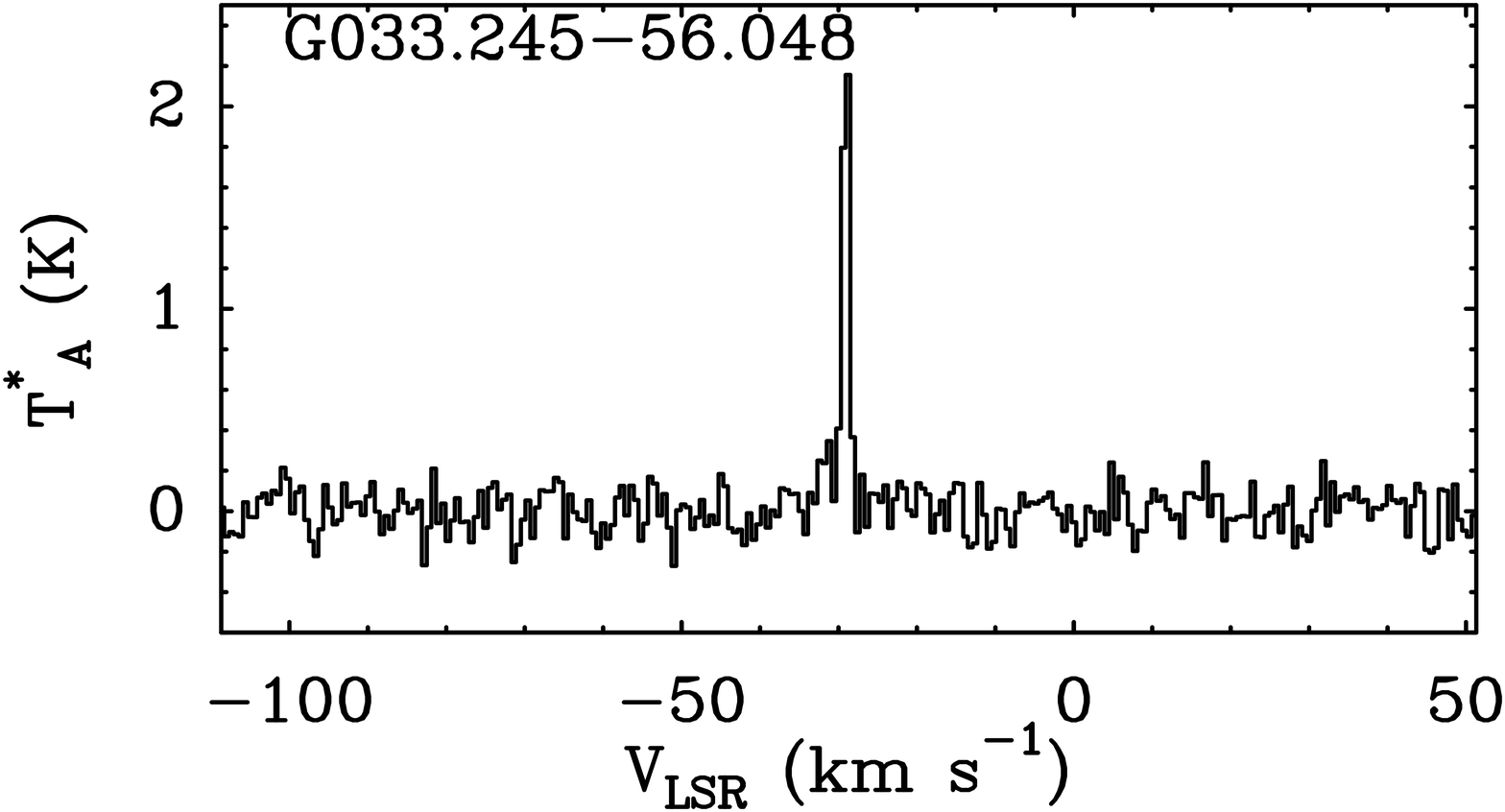}
\includegraphics[width=5.0cm]{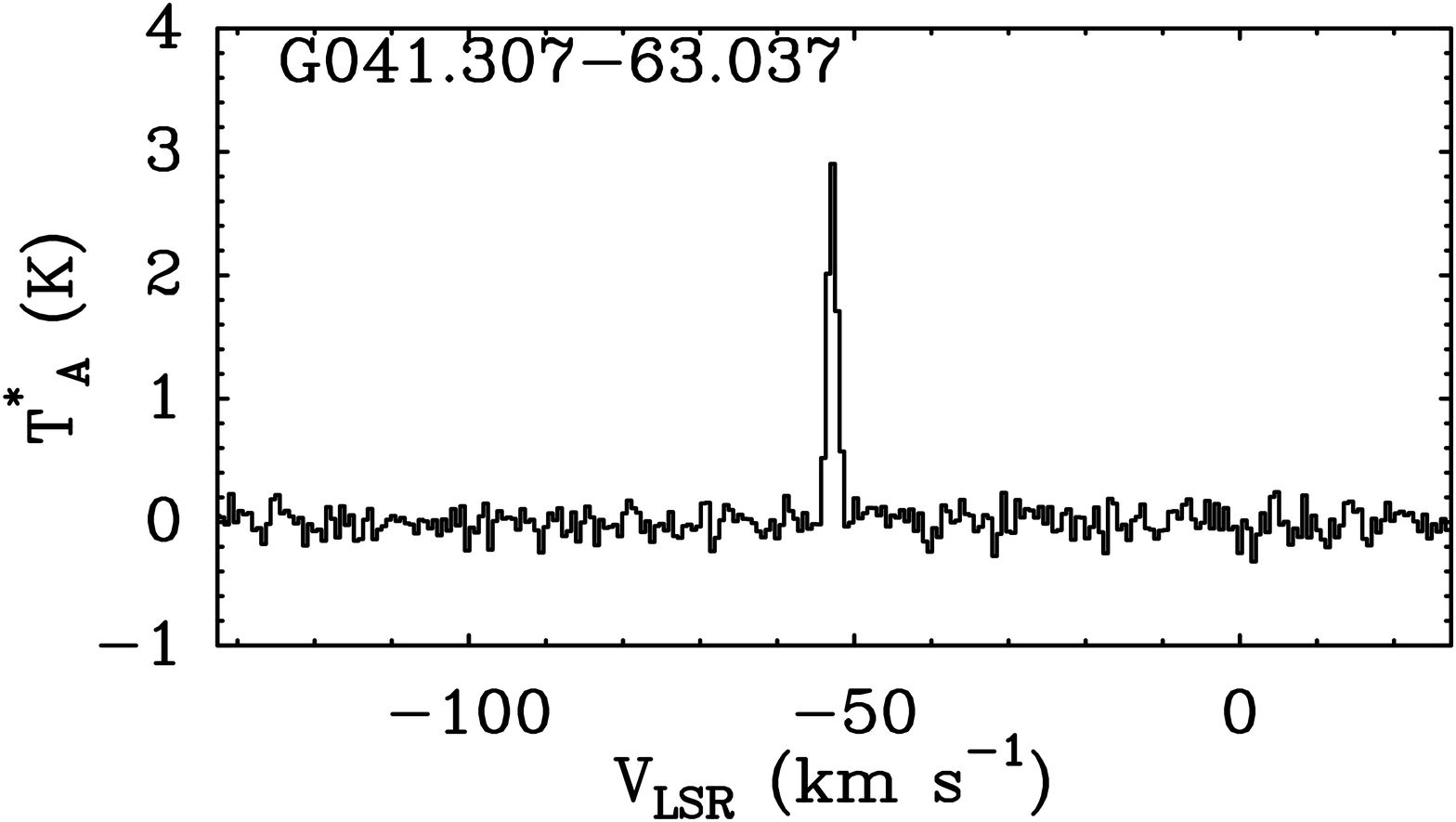}
\includegraphics[width=5.0cm]{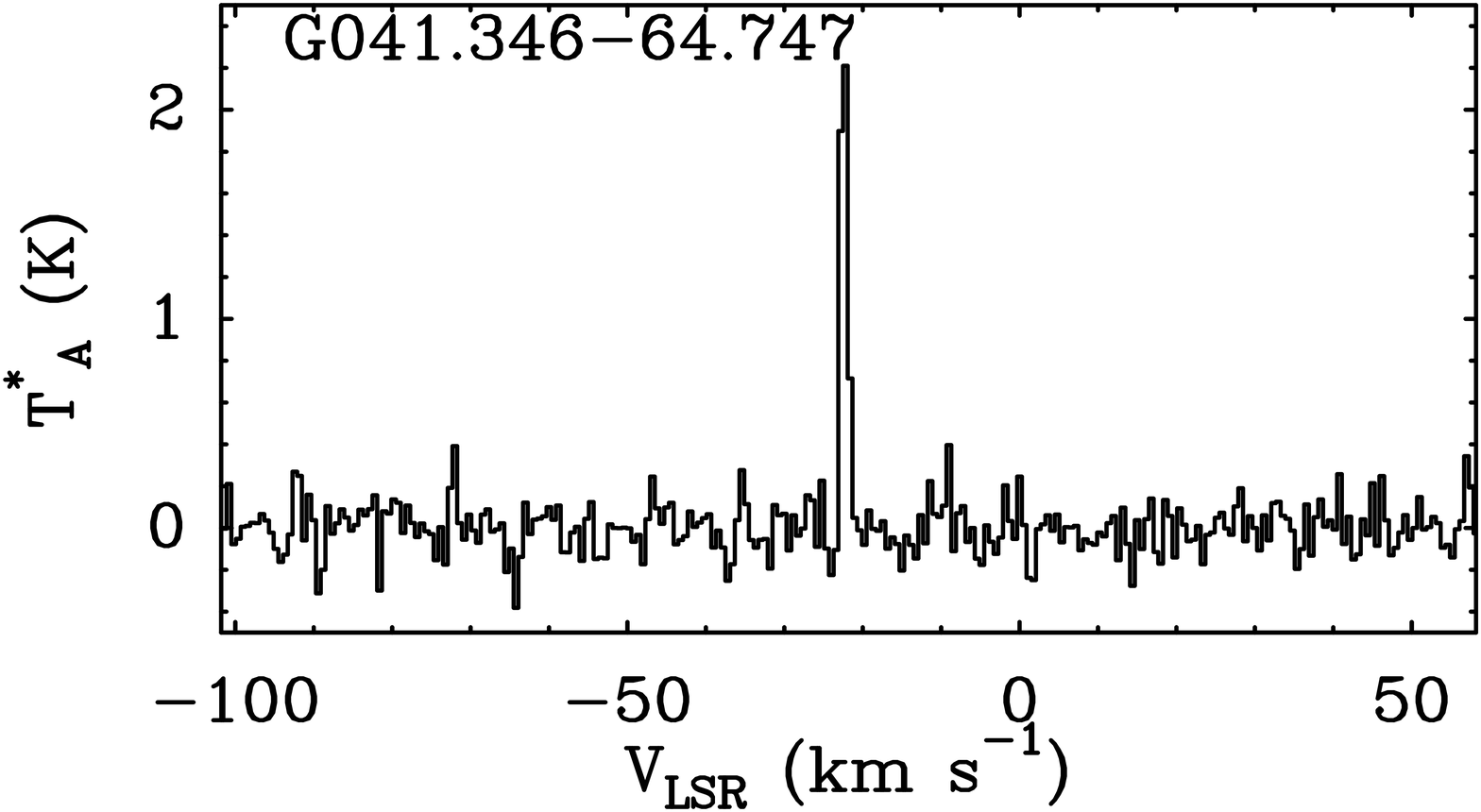} \\
\includegraphics[width=5.0cm]{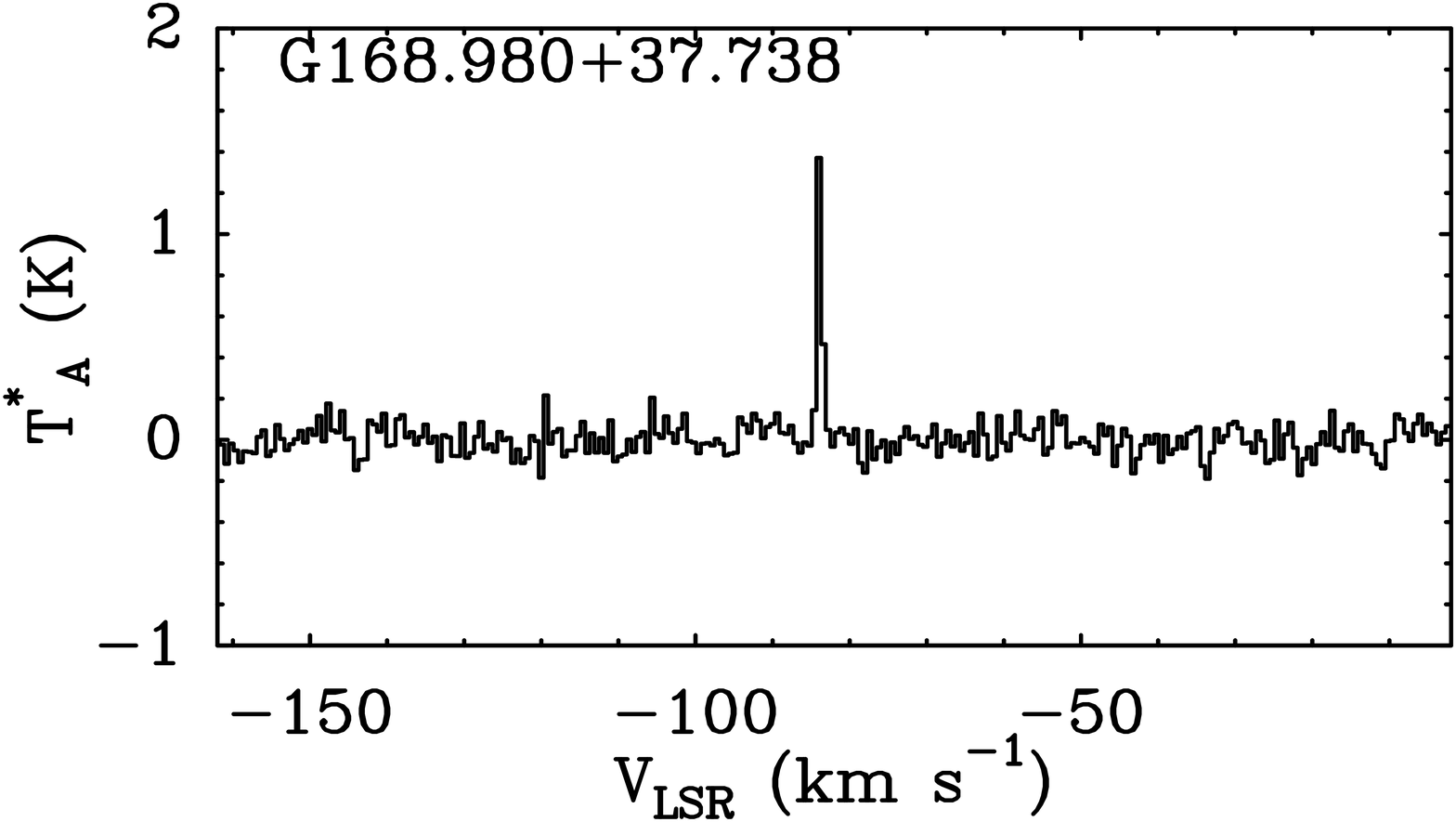}
\includegraphics[width=5.0cm]{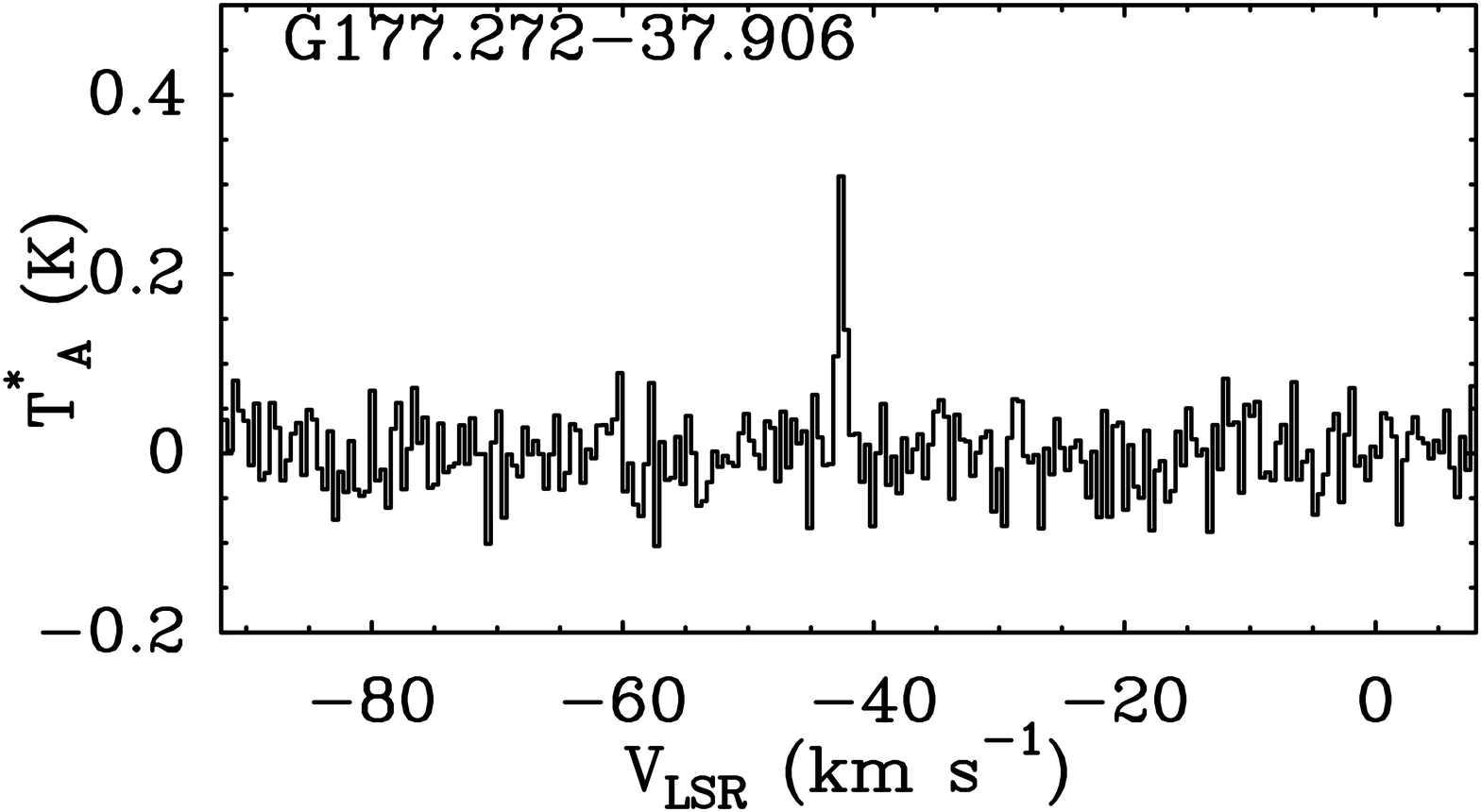}   
\includegraphics[width=5.0cm]{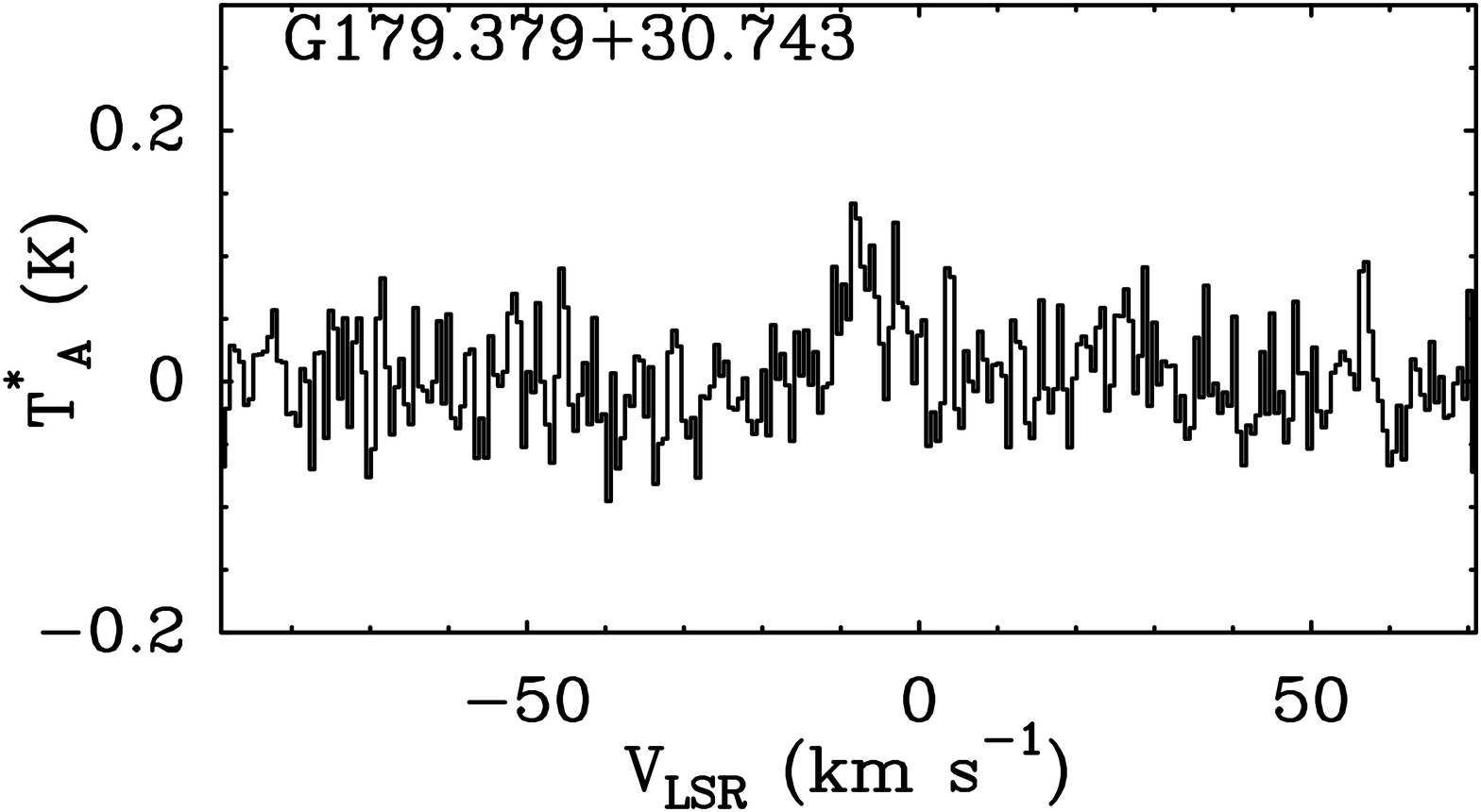} \\
\includegraphics[width=5.0cm]{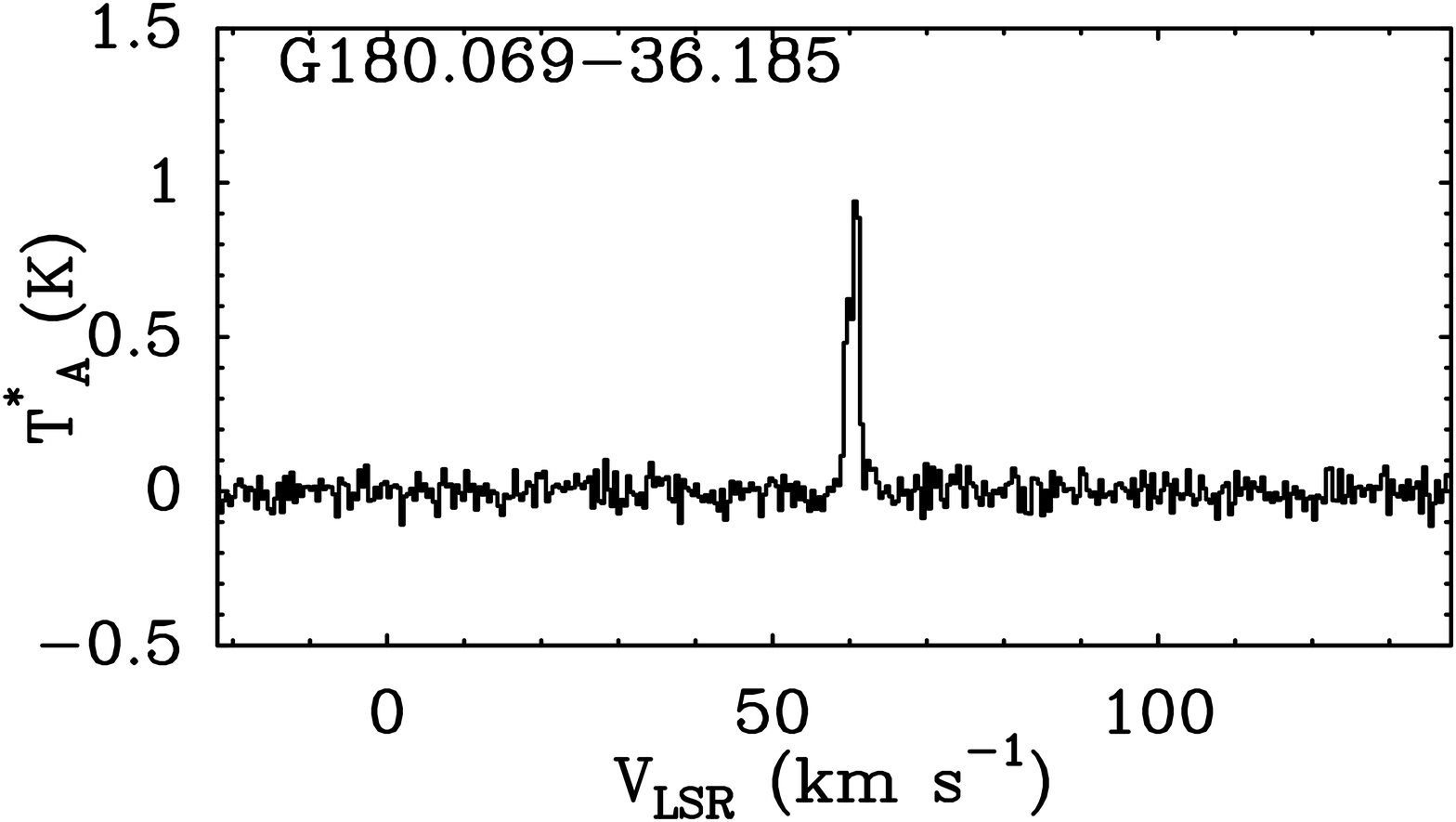}   
\includegraphics[width=5.0cm]{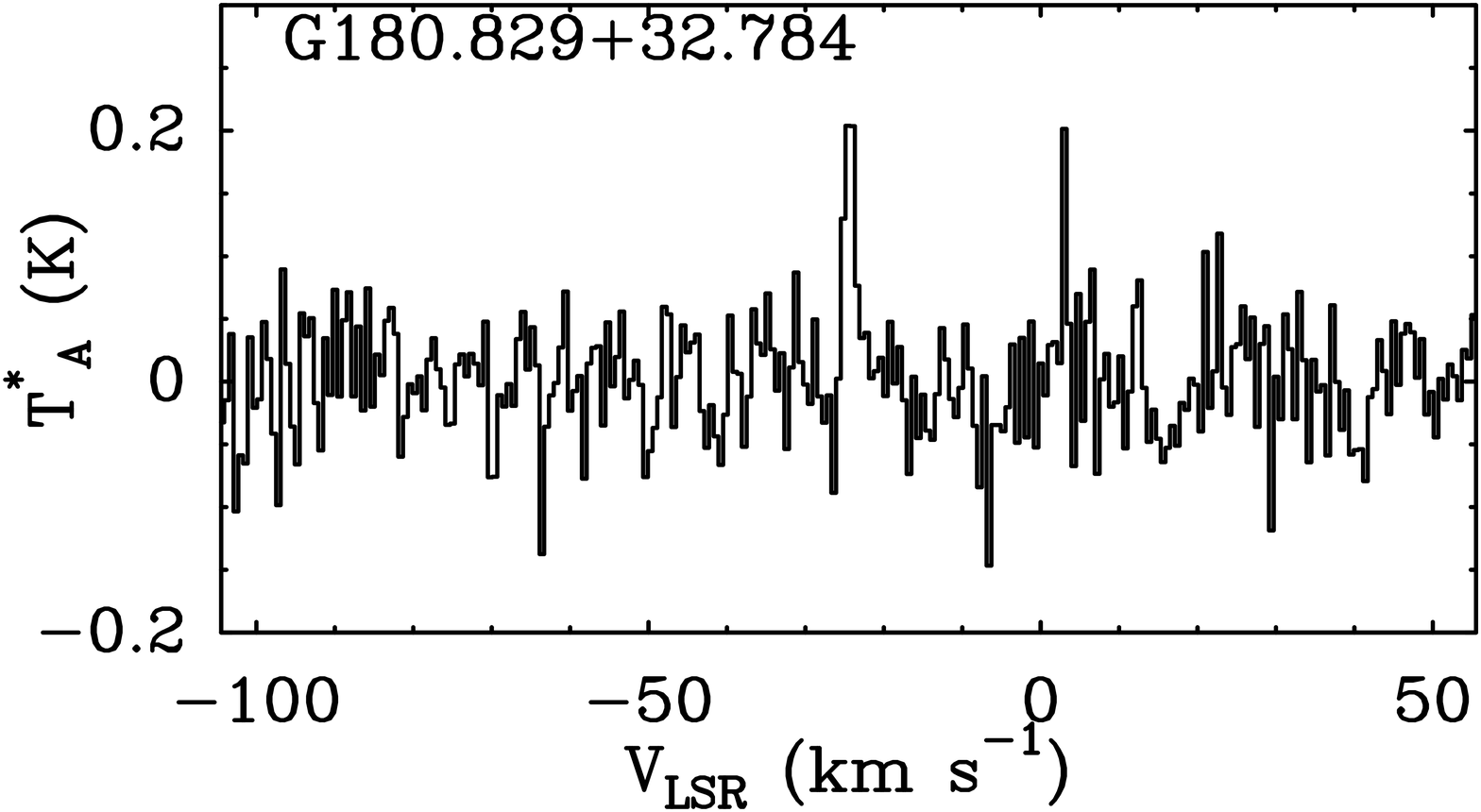}
\includegraphics[width=5.0cm]{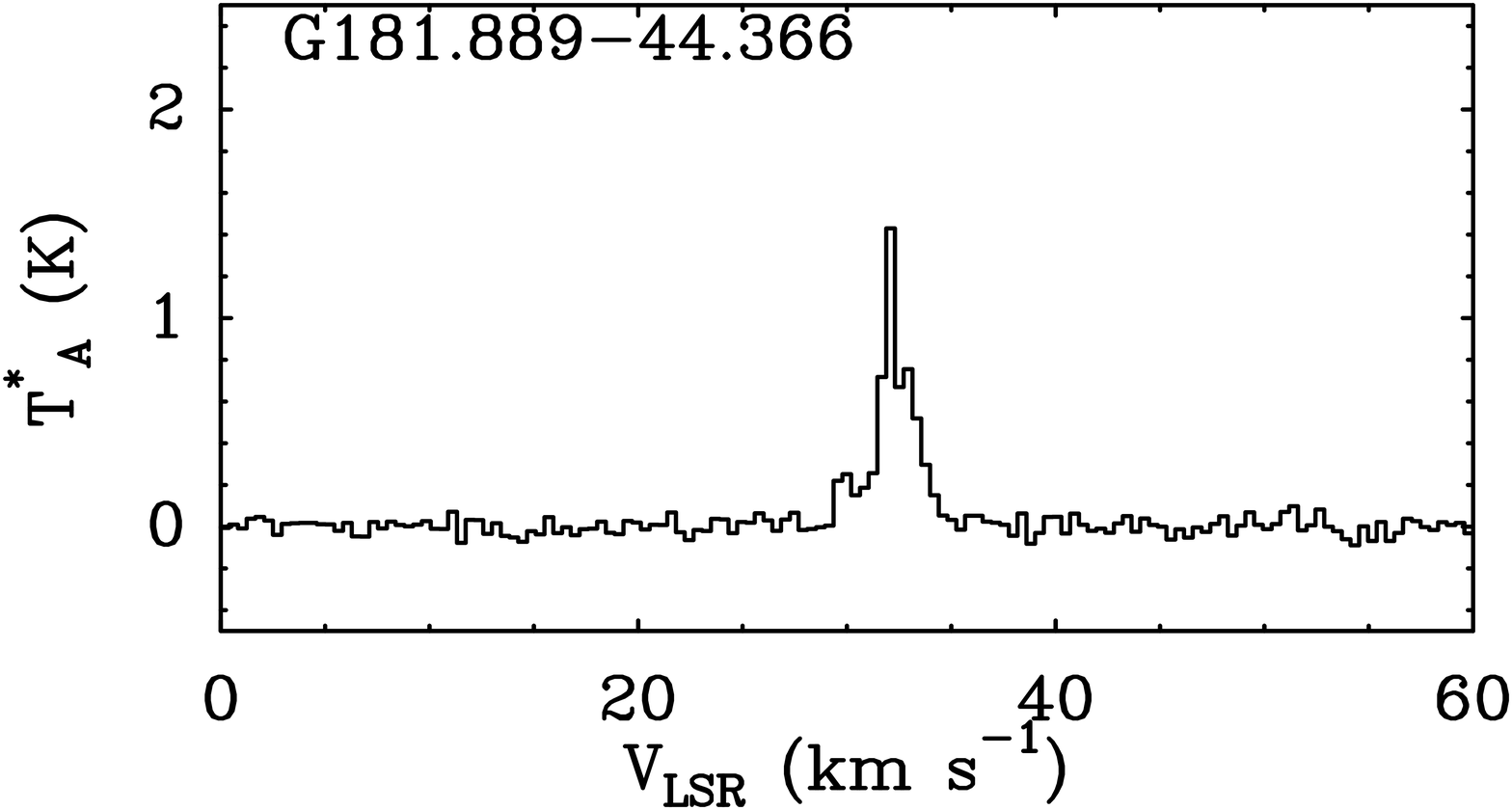} \\
\includegraphics[width=5.0cm]{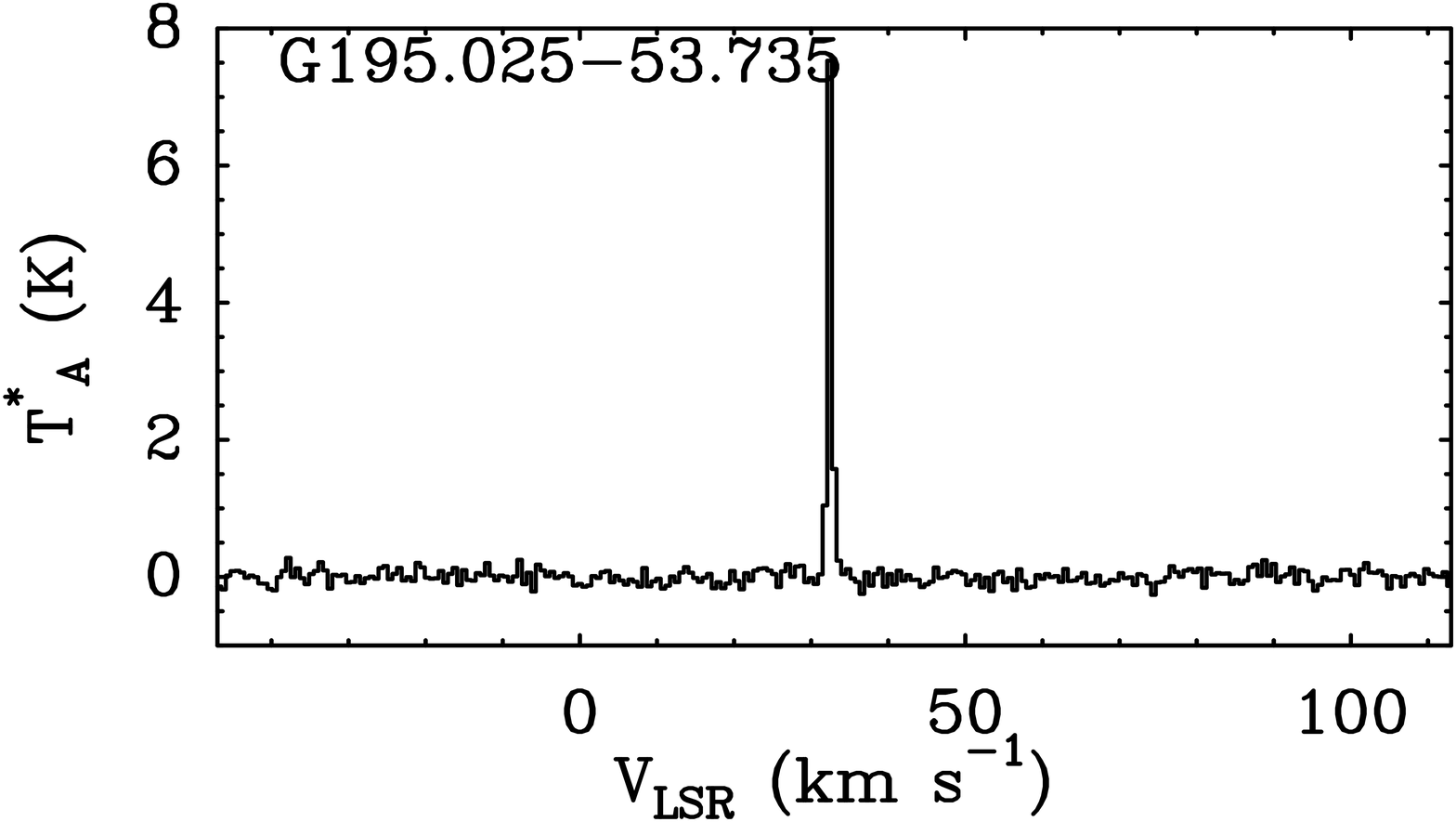}
\includegraphics[width=5.0cm]{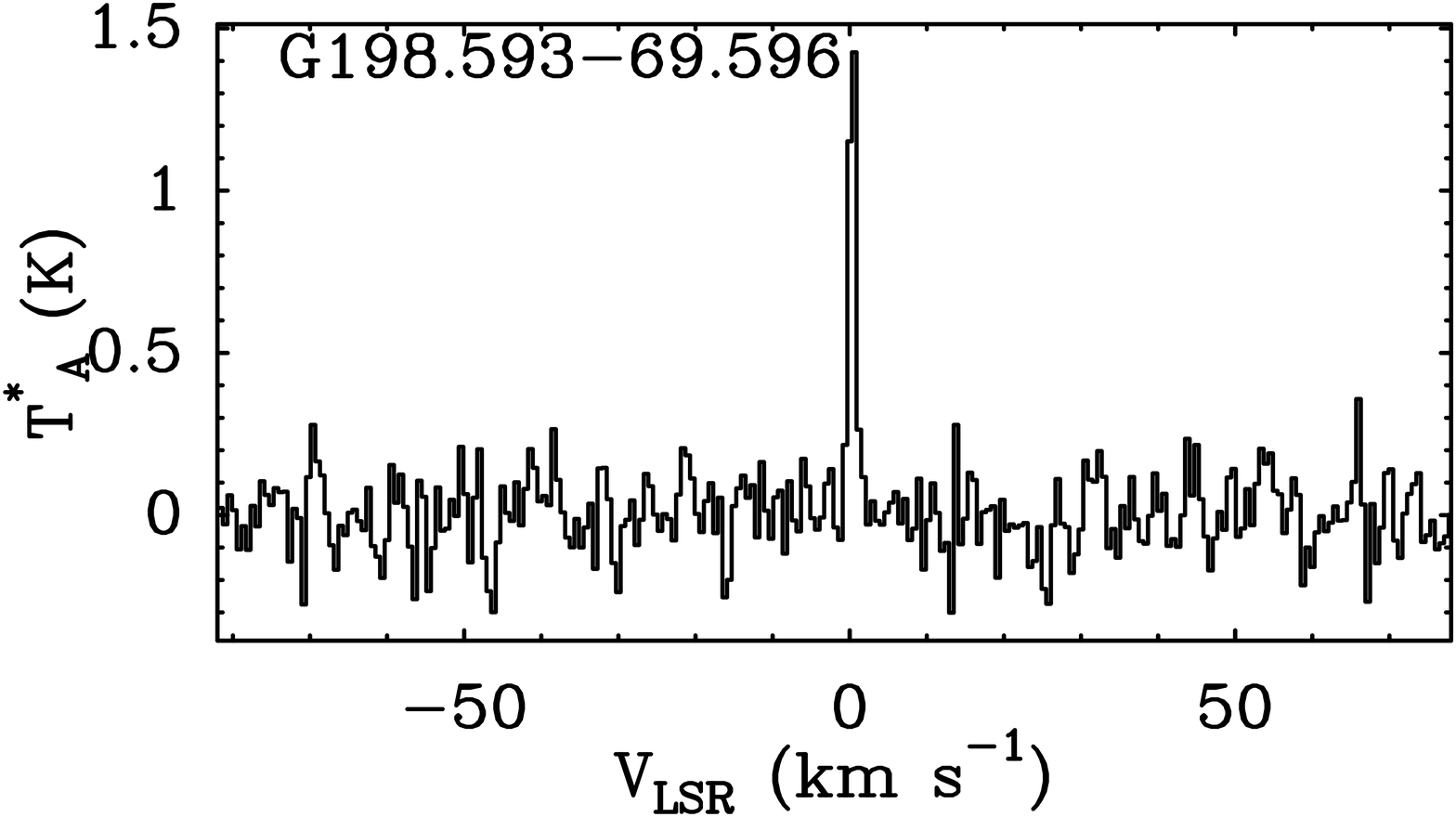}
\includegraphics[width=5.0cm]{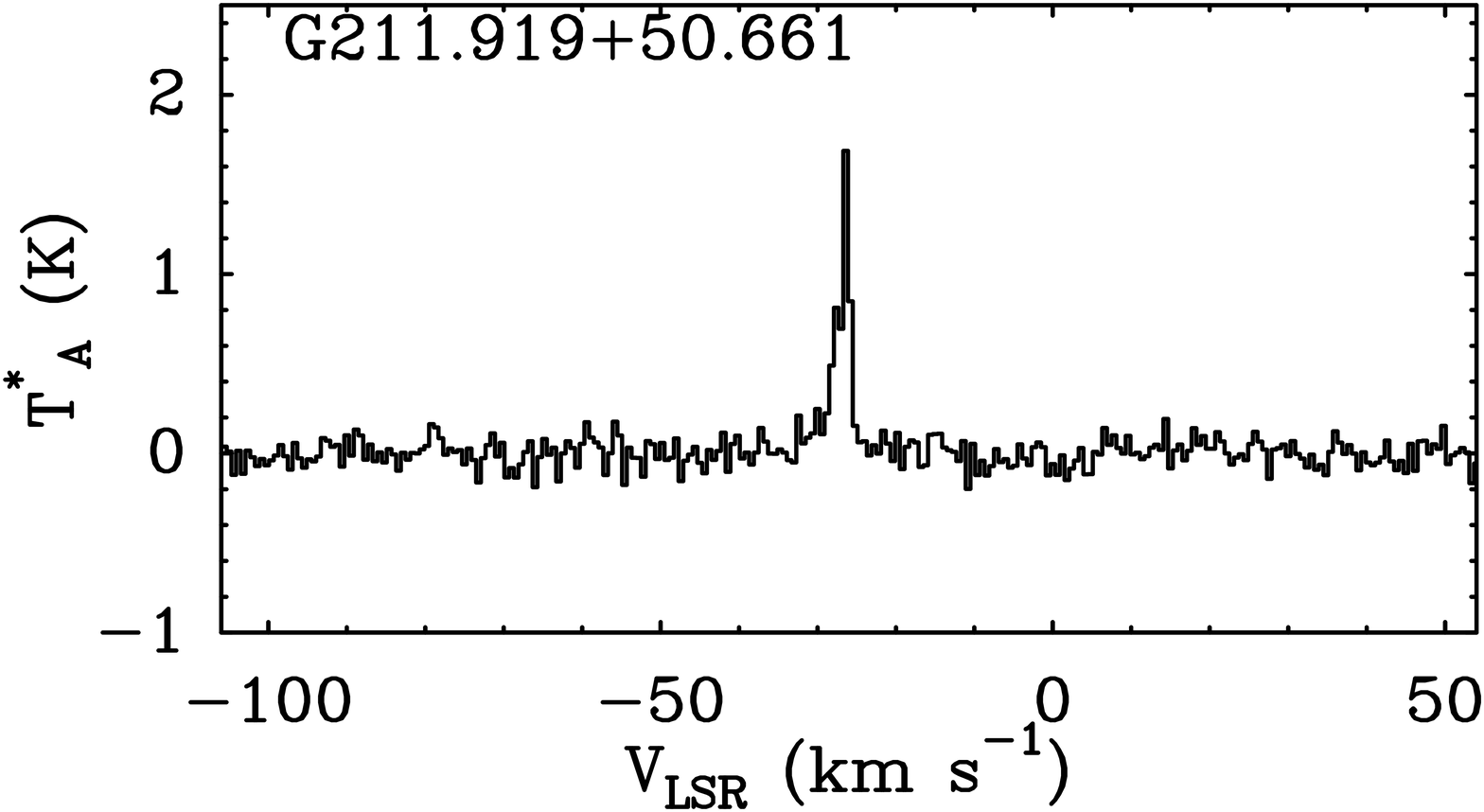} \\
\includegraphics[width=5.0cm]{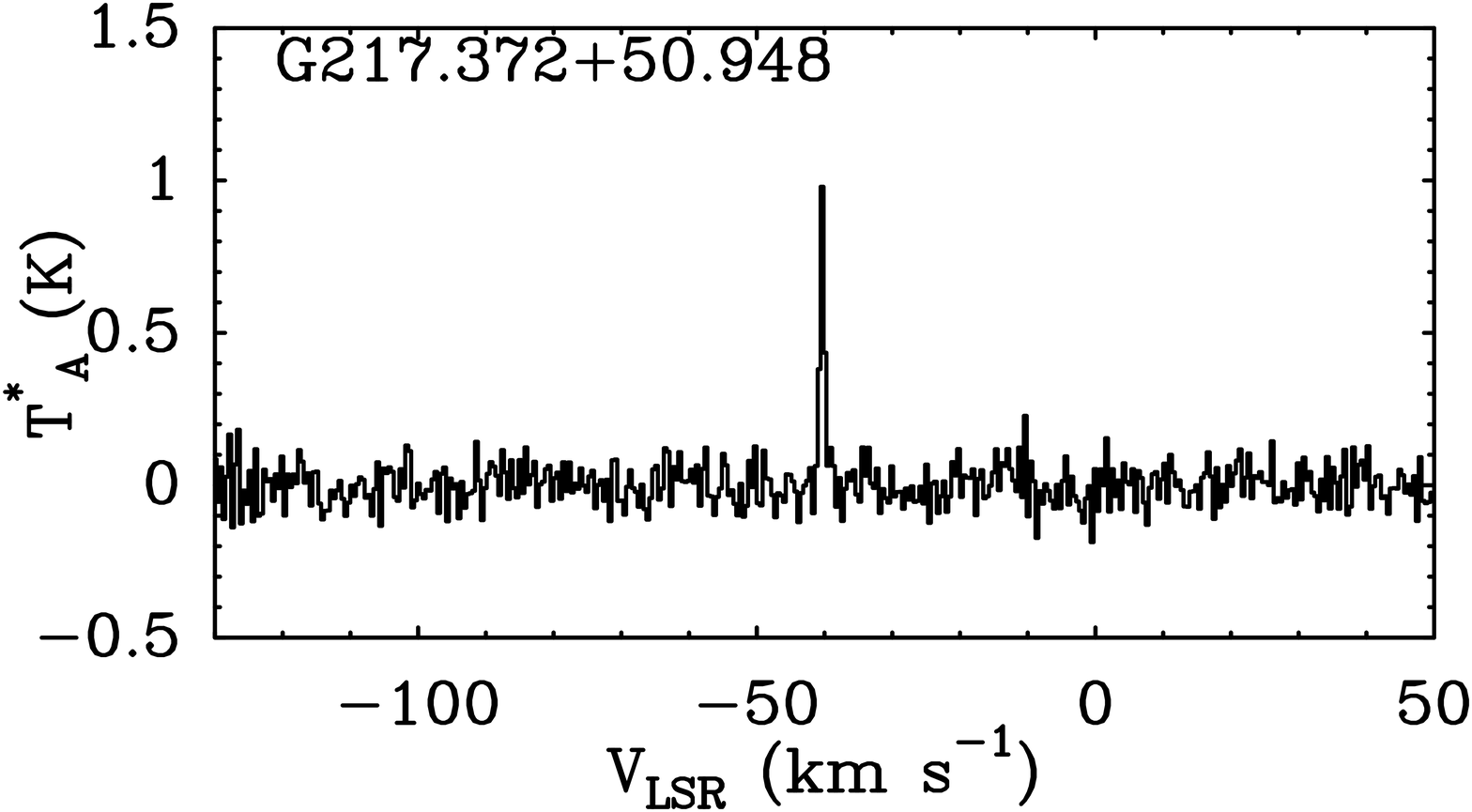}
\includegraphics[width=5.0cm]{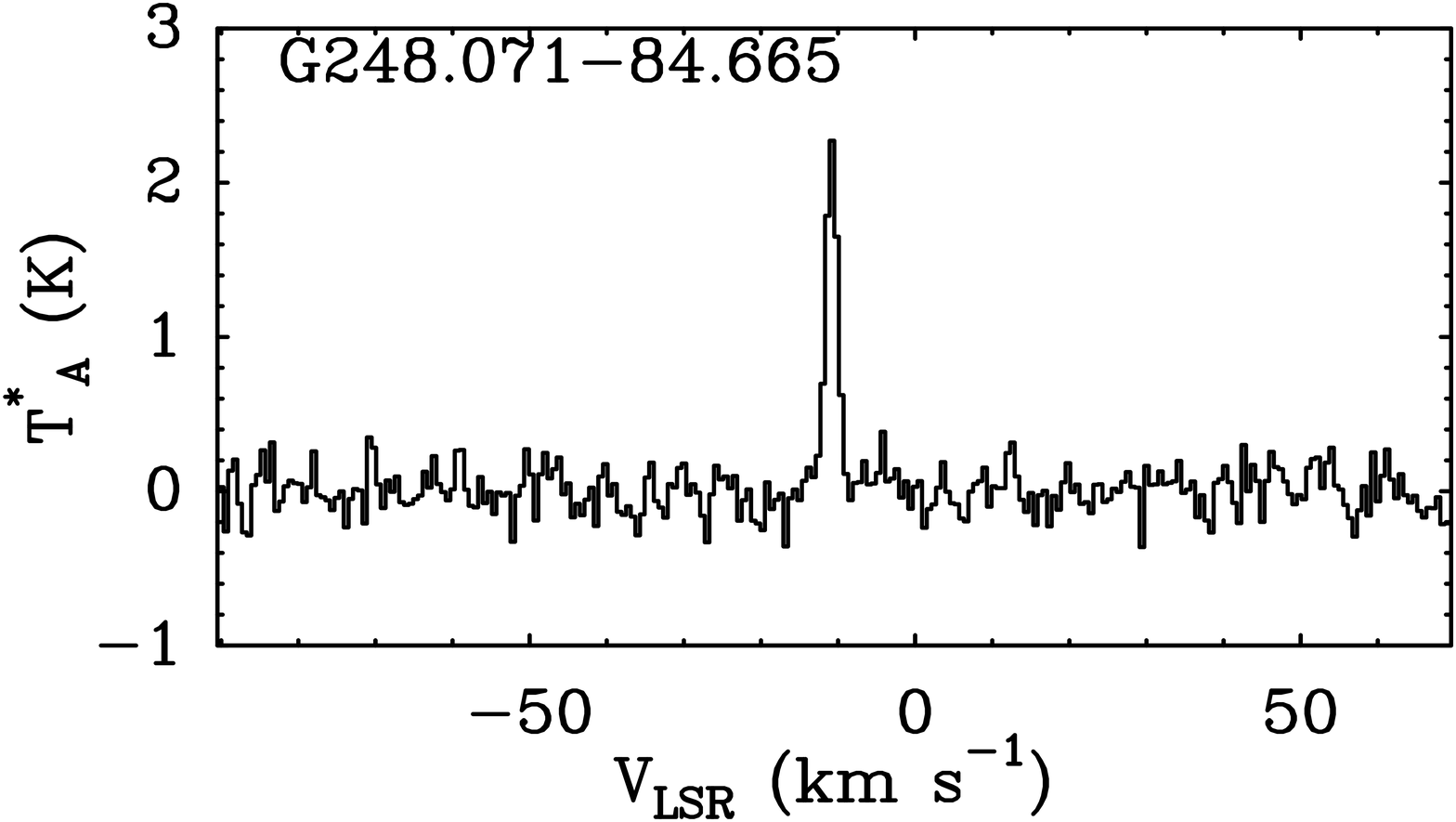}
\includegraphics[width=5.0cm]{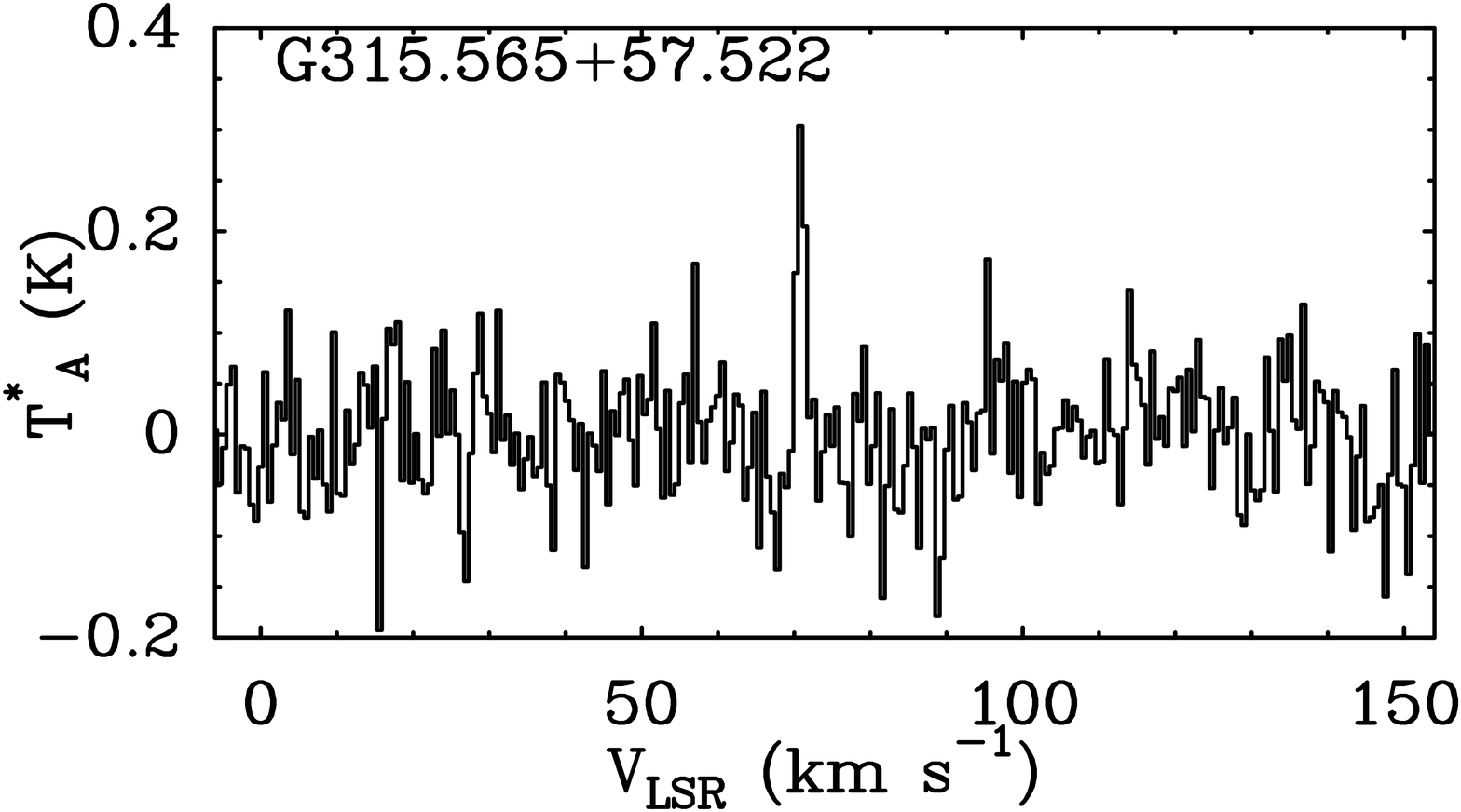} \\
\includegraphics[width=5.0cm]{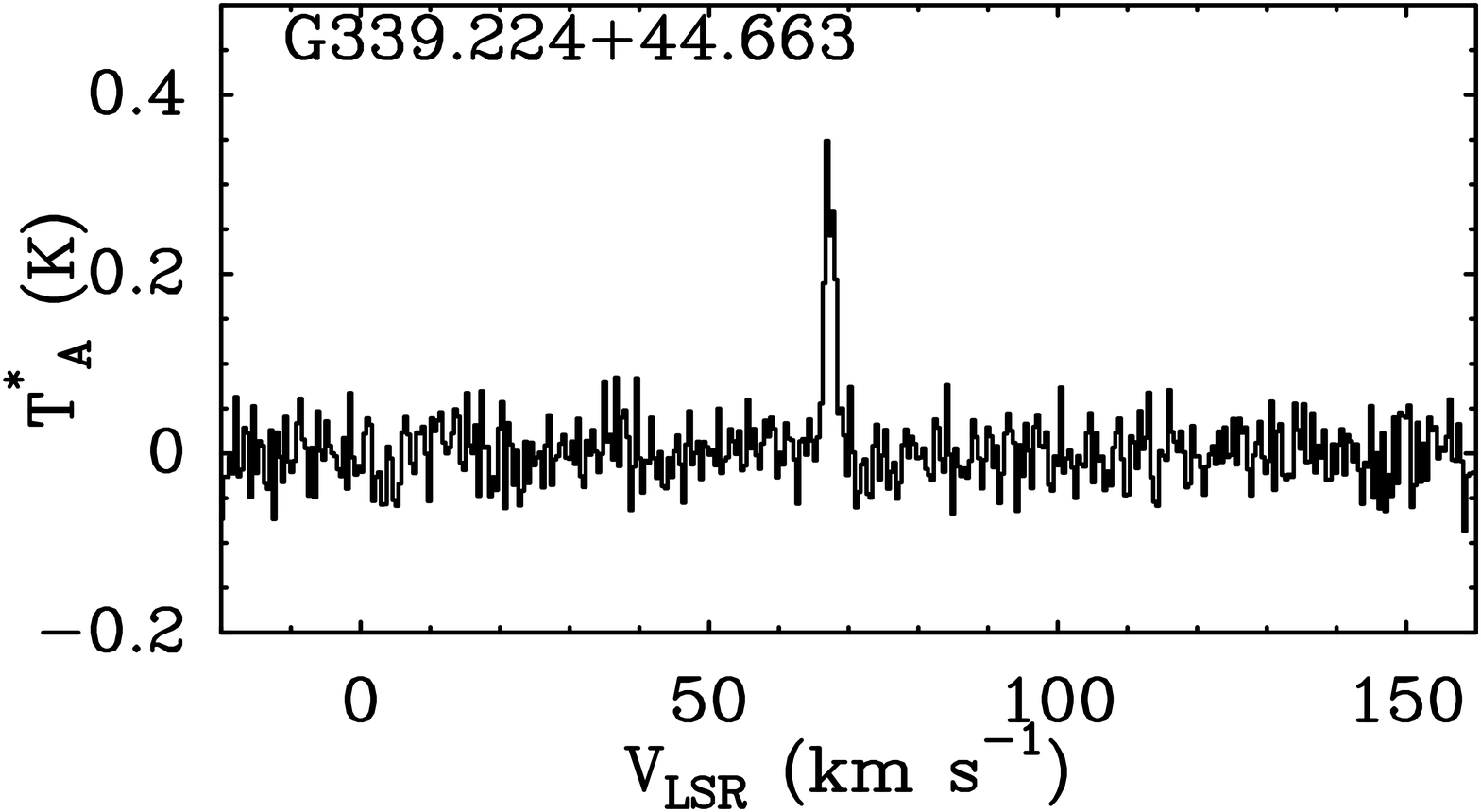}
\includegraphics[width=5.0cm]{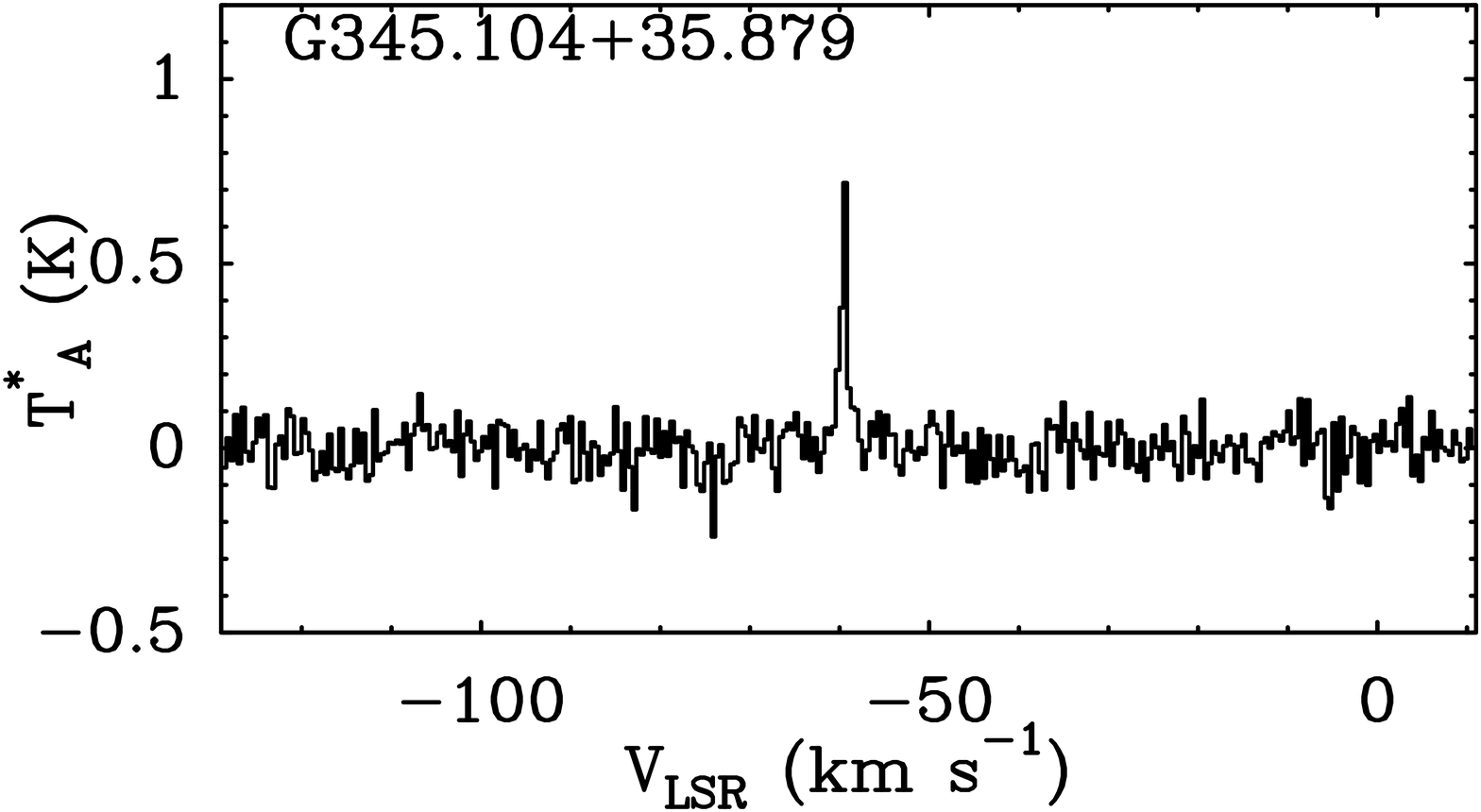}
\includegraphics[width=5.0cm]{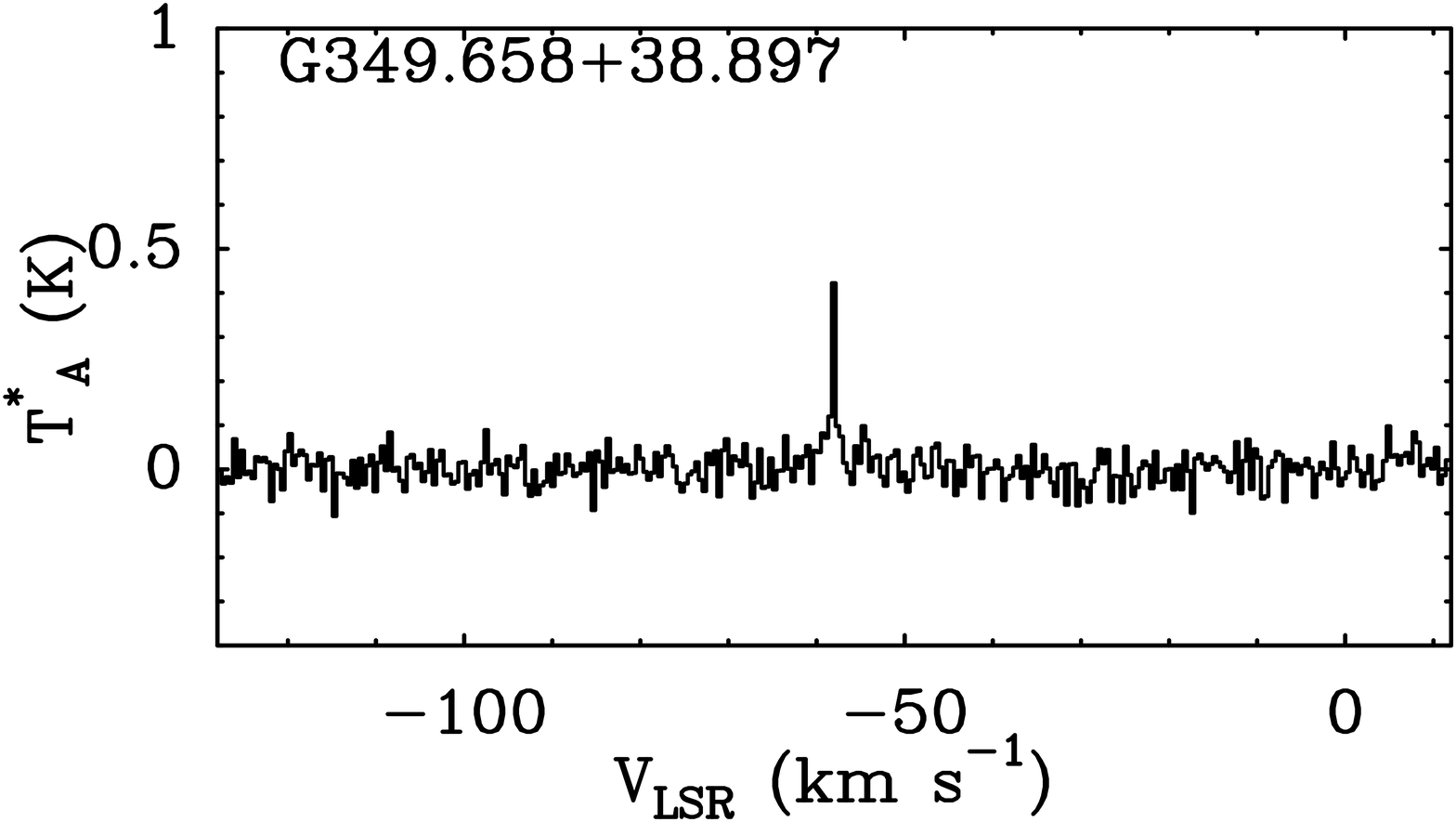} \\
%Fig.~\ref{fig-A3} continue
\caption{The 22 GHz H$_2$O maser spectra. Source names are noted on individual spectra. \label{fig-3}}
\end{figure*}

The maser spectra are presented in Figure \ref{fig-3}. Within these 21 H$_2$O
maser sources listed in Table \ref{tab-1}, 14 sources are detected by Nobeyama
45m telescope and 7 sources are detected by Tidbinbilla 70m telescope (denoted
in column 13 of Table \ref{tab-1}). In addition, SiO maser survey have been
conducted for 17 of these sources, which yielded 12 detection and 5 non
detection of SiO maser \citep{2018MNRAS.473.3325W}. 

%There are 4 sources, G181.889$-$44.366, G177.272$-$37.906, G217.372+50.948,
%G349.658+38.897 have detecion of H$_2$O masers with Tidbinbilla 70m telescope,
%that are still not covered by any SiO maser survey. Taking into account of
%previous 

\section{Discussions} \label{sec:discuss}
Combining the H$_2$O maser survey with the previous SiO maser survey results, we
formed a sample of 52 maser-traced O-rich AGBs towards the Sgr orbital plane.
In this section, we investigate the 3D distribution and 3D kinematics of this
sample with latest Gaia astrometric results \citep{2021A+A...649A...1G},
parallaxes and proper motions with Gaia Early Data Release 3 (Gaia EDR3) data.

\begin{figure*}
%\figurenum{3}
\includegraphics[width=15cm]{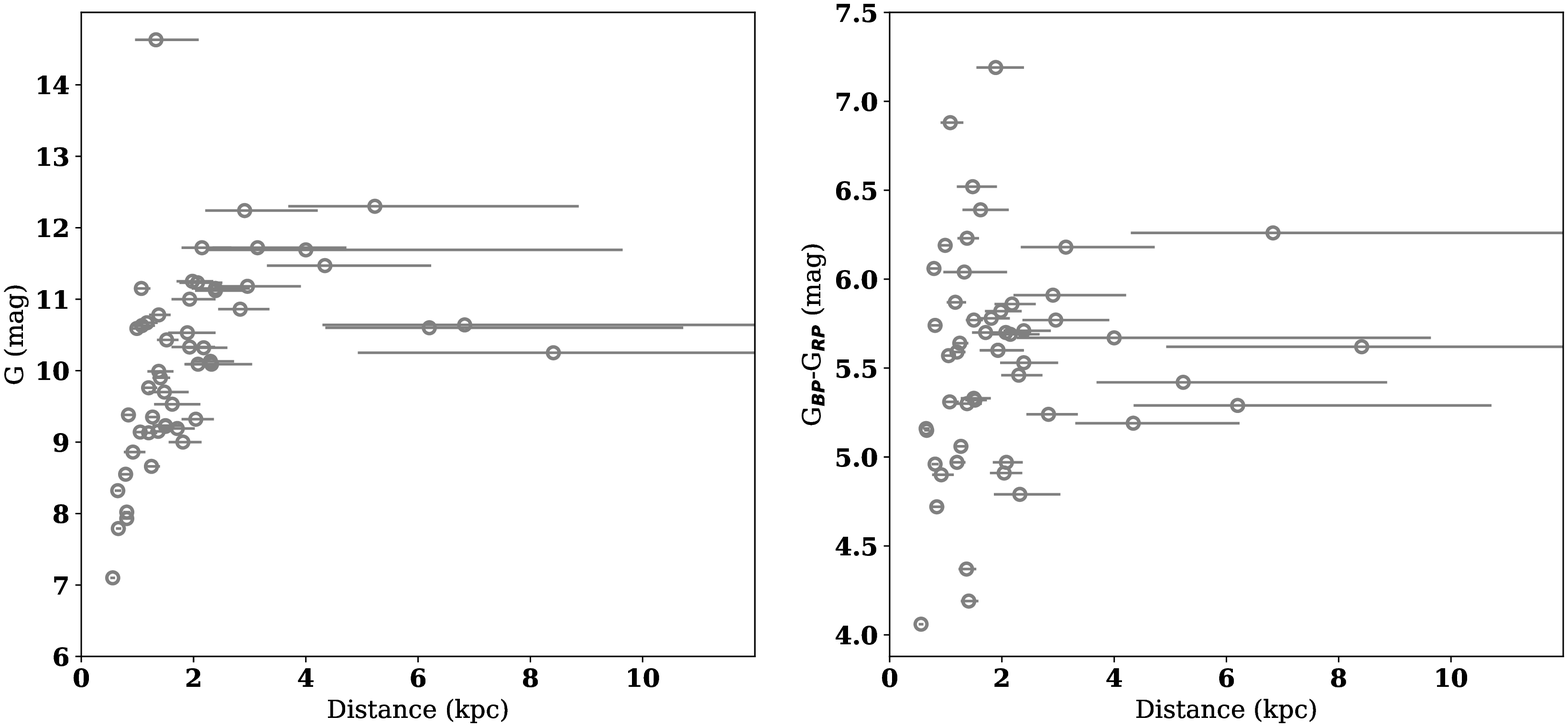}
\caption{
Gaia EDR3 parallax distances versus mean G band magnitudes (\textit{Left
panel}), and G$_{BP}$-G$_{RP}$ colors (\textit{Right panel}). The horizontal error
bars denote uncertainties of distances.
\label{fig-4}}
\end{figure*}

\subsection{Distance and Galactic distribution}\label{subsec:distance}

Accurate distances are the key parameters to investigate the distribution of
Galactic objects \citep{2012PASJ...64..136H, 2014ApJ...783..130R}. In our
previous investigations Period Luminosity Relation (PLR) distances and WISE
luminosity distances are used, with distance uncertainties larger than 30\% or
even higher.  \citet{2018A+A...618A..58M} studied properties of long-period
variables using the Gaia DR2 data and found that parallaxes uncertainties can
reach close to 10\% for LPVs with G$_{BP} -$~G$_{RP}$<=2.5~mag, and increases
to 100\% for LPVs with G$_{BP} -$~G$_{RP}$>5~mag. Recently, Gaia EDR3 data are
released, the parallax precision are increased by 30\%
\citep{2021A+A...649A...1G}. In Table \ref{tab-2}, we compiled tables including
the parallaxes, proper motions of these maser-traced AGBs cross matched from
Gaia EDR3 data. In Figure \ref{fig-4} we shows diagrams of distance versus G
band magnitude (left panel) and distance versus G$_{BP} -$~G$_{RP}$ color of
Table \ref{tab-2} sources.  The G$_{BP} -$~G$_{RP}$ colors of this maser-traced
AGB stars are within 4 to 7~mag, parallaxes uncertainties for sources within
3~kpc are all within 5\%-30\%.  While for sources with distances larger than
4~kpc (source ID 1, 17, 22, 43, 44, 47), parallaxes uncertainties can be reach
up to 30\%-70\%, with distance uncertainties larger than 2~kpc. In the left
panel of Figure \ref{fig-4}, it can also been seen a statistically linear
relation between G band magnitude and distances. This can be due to the reason
that most of sources detected in our survey are Miras with periods within 230
to 330 days, therefore, has a similar luminosities, the linear relation
indicates that, statistically, faint sources are also sources with large
distances. Apart from distances and proper motions, the maser radio V$_{LSR}$
velocities are also compiled in the last column of Table \ref{tab-2}.

\begin{figure*}
%\figurenum{3}
\includegraphics[width=15cm]{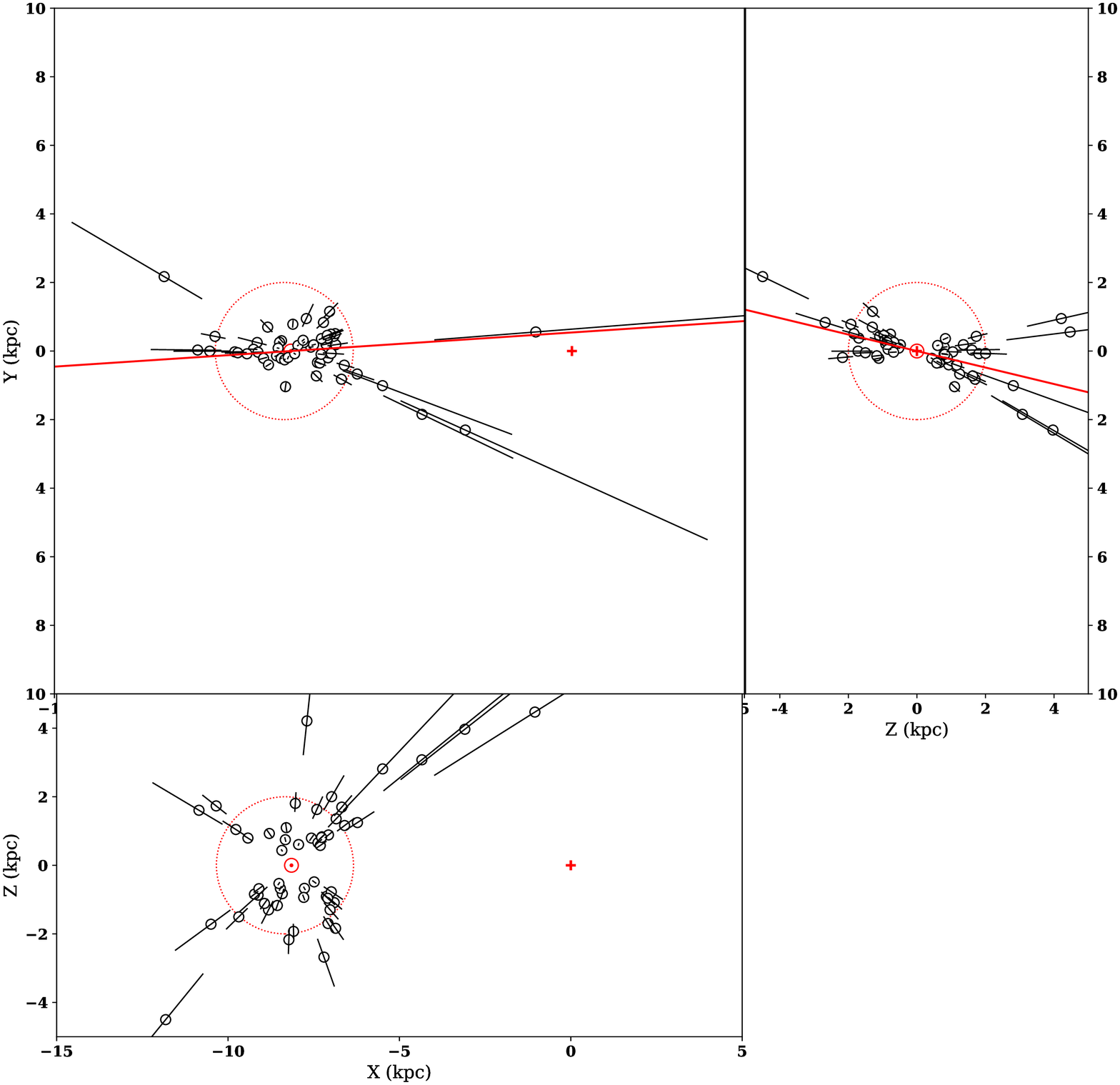}
\caption{
3D locations of SiO and H$_2$O masers in Galactocentric Cartesian coordinates,
with the Galactic Centre at the zero-point and the Sun at [-8.15, 0.0]~kpc.
The normal direction of the Sgr orbital plane is towards ($l$, $b$) =
(273$^\circ$.8, $-$13$^\circ$.5), which is tipped by about 77$^\circ$ with
respect to the Galactic plane, with Sgr dSph located at a distance of
$\sim$20~kpc \citep{2004AJ....127.2031K} in the direction of ($l$, $b$) =
(5$^\circ$.608, $-$14$^\circ$.086). Thus the separation angle between the X-Z
plane and the Sgr orbital plane is very small (14$^\circ$), in the X-Y and Z-Y
plots, the red lines are the intersection between the Sgr orbital plane and
X-Y and Y-Z plane. The dotted red circle denotes a 2~kpc circle around the
Sun.\label{fig-5}}
\end{figure*}

In Figure \ref{fig-5}, we show the 3D locations of the maser-traced AGB sample
listed in Table \ref{tab-2}. From the figure, it can be clearly seen a linearly
elongated structure traced by sources (Source ID within 40 to 51 in Table
\ref{tab-2}) towards the ($l$, $b$)~$\sim$~(340, 40) direction, the
constellation of Libra. It looks like that the elongated structure cross the
Galactic plane with the intersection within $\sim$2~kpc around the Sun and
even extended to the southern hemispheric source.

\cite{2004AJ....128..245M} studied the properties of Sagittarius Stellar Stream
with M giants samples using 2MASS all sky survey data and found a breadth of
8$-$10~kpc, all sky M giants distributions may suggest the Sgr debris from the
leading arm to be falling down onto the Galactic plane with an apparently near
`direct hit' on the solar neighborhood \citep{2004PASA...21..197M}. Further we
found the X-Z plane projection of these maser-traced AGB stars shown in Figure
\ref{fig-5} is very closed to the L2 wraps of \citet{2010ApJ...714..229L}. To
verify whether these sources are tidal debis of Sagittarius Stellar Stream or
just foreground thick disc stars, we further studied their kinematics. 

\begin{table*}
%\footnotesize
%\setlength\tabcolsep{1pt}
\renewcommand\arraystretch{1.23}
\caption{Gaia EDR3 Parallaxes and Proper motions of O-rich AGBs with detection of H$_2$O and/or SiO masers.
\label{tab-2}}
\begin{tabular}{ccrcccrrr}
\hline
ID & Source & G$\quad$ & G$_{BP}$-G$_{RP}$  & Parallax & Distance & $\mu_x\quad\quad$ &  $\mu_y\quad\quad$ & $V_{LSR}\quad$ \\
   & Name   & (mag)  & (mag)              & (mas)    & (kpc)    & (mas yr$^{-1}$) & (mas yr$^{-1}$)      & (km~s$^{-1}$) \\
\hline
  1 & G004.482$+$32.104  &10.25 & 5.62 & 0.119 $\pm$ 0.083  & 8.41$^{+19.5}_{-3.46}$&  -4.029 $\pm$ 0.090 & -2.629 $\pm$ 0.058 &  -8.8  \\
  2 & G008.104$+$45.840  &10.53 & 7.19 & 0.529 $\pm$ 0.107  & 1.89$^{+0.48}_{-0.32}$& -14.224 $\pm$ 0.131 &  -8.84 $\pm$ 0.116 &  41.0  \\
  3 & G011.025$+$53.268  &10.59 & 6.19 & 1.014 $\pm$ 0.091  & 0.99$^{+0.10}_{-0.08}$&   7.696 $\pm$ 0.098 & -0.175 $\pm$ 0.089 &   4.5  \\
  4 & G011.159$-$41.196  & 9.99 & 5.30 & 0.725 $\pm$ 0.106  & 1.38$^{+0.24}_{-0.18}$&  -6.902 $\pm$ 0.093 &-16.568 $\pm$ 0.079 &  16.2  \\
  5 & G015.405$-$35.139  & 9.38 & 4.72 & 1.184 $\pm$ 0.077  & 0.84$^{+0.06}_{-0.05}$&  -11.08 $\pm$ 0.077 &  0.010 $\pm$ 0.059 &  19.9  \\
  6 & G019.002$-$39.495  & 9.19 & 5.70 & 0.584 $\pm$ 0.085  & 1.71$^{+0.29}_{-0.22}$&   8.111 $\pm$ 0.089 &  0.777 $\pm$ 0.073 & -51.8  \\
  7 & G019.509$-$56.308  & 9.32 & 4.91 & 0.490 $\pm$ 0.064  & 2.04$^{+0.30}_{-0.23}$&  -0.902 $\pm$ 0.061 & -0.762 $\pm$ 0.058 & -29.0  \\
  8 & G021.513$-$53.023  &10.13 & 5.46 & 0.435 $\pm$ 0.064  & 2.30$^{+0.40}_{-0.29}$&  10.641 $\pm$ 0.063 &-14.670 $\pm$ 0.051 & -98.0  \\
  9 & G022.158$+$40.858  & 9.35 & 5.06 & 0.785 $\pm$ 0.056  & 1.27$^{+0.10}_{-0.09}$&  -8.926 $\pm$ 0.049 &  8.041 $\pm$ 0.041 & -15.0  \\
 10 & G022.943$-$31.448  & 9.70 & 6.52 & 0.674 $\pm$ 0.145  & 1.48$^{+0.41}_{-0.26}$&  10.203 $\pm$ 0.134 &-19.248 $\pm$ 0.104 &   9.8  \\
 11 & G023.376$-$39.816  & 9.23 & 5.33 & 0.668 $\pm$ 0.106  & 1.50$^{+0.28}_{-0.21}$&  -5.073 $\pm$ 0.096 & -0.771 $\pm$ 0.070 & -23.4  \\
 12 & G033.245$-$56.048  & 8.02 & 4.96 & 1.239 $\pm$ 0.064  & 0.81$^{+0.04}_{-0.04}$&  31.843 $\pm$ 0.060 & -6.247 $\pm$ 0.054 & -29.1  \\
 13 & G038.070$+$66.469  & 7.79 & 5.15 & 1.520 $\pm$ 0.059  & 0.66$^{+0.03}_{-0.02}$& -25.537 $\pm$ 0.051 & 11.968 $\pm$ 0.060 & -42.4  \\
 14 & G041.307$-$63.037  & 9.14 & 5.57 & 0.949 $\pm$ 0.069  & 1.05$^{+0.08}_{-0.07}$&   0.774 $\pm$ 0.078 &-19.787 $\pm$ 0.066 & -52.8  \\
 15 & G041.346$-$64.747  &11.18 & 5.77 & 0.338 $\pm$ 0.081  & 2.96$^{+0.93}_{-0.57}$&   7.456 $\pm$ 0.088 & -5.026 $\pm$ 0.077 & -22.4  \\
 16 & G045.734$-$38.770  &11.23 & 5.70 & 0.482 $\pm$ 0.081  & 2.07$^{+0.42}_{-0.30}$&  -2.832 $\pm$ 0.091 & -5.103 $\pm$ 0.068 &  26.3  \\
 17 & G064.549$+$76.014  &11.47 & 5.19 & 0.230 $\pm$ 0.069  & 4.34$^{+1.87}_{-1.01}$&  -9.971 $\pm$ 0.047 & -2.271 $\pm$ 0.053 &  26.3  \\
 18 & G085.581$-$67.859  &10.09 & 4.97 & 0.481 $\pm$ 0.056  & 2.08$^{+0.27}_{-0.22}$&   3.098 $\pm$ 0.058 & -5.951 $\pm$ 0.039 &  50.9  \\
 19 & G131.720$-$64.091  & 8.86 & 4.90 & 1.089 $\pm$ 0.198  & 0.92$^{+0.20}_{-0.14}$& -11.187 $\pm$ 0.212 &  3.084 $\pm$ 0.161 &   4.0  \\
 20 & G133.797$-$53.388  & 9.53 & 6.39 & 0.619 $\pm$ 0.142  & 1.62$^{+0.48}_{-0.30}$&   0.669 $\pm$ 0.153 & -3.042 $\pm$ 0.134 &   4.3  \\
 21 & G141.940$-$58.536  & 8.55 & 6.06 & 1.266 $\pm$ 0.095  & 0.79$^{+0.06}_{-0.06}$&   4.792 $\pm$ 0.116 &-12.788 $\pm$ 0.061 & -57.0  \\
 22 & G149.396$-$46.550  &10.60 & 5.29 & 0.161 $\pm$ 0.068  & 6.20$^{+4.50}_{-1.83}$&   2.723 $\pm$ 0.081 & -1.552 $\pm$ 0.058 &   9.5  \\
 23 & G165.616$-$40.899  &14.63 & 6.04 & 0.751 $\pm$ 0.268  & 1.33$^{+0.74}_{-0.35}$&   4.485 $\pm$ 0.303 & -7.364 $\pm$ 0.301 &  13.0  \\
 24 & G166.965$-$54.751  & 8.32 & 5.16 & 1.537 $\pm$ 0.080  & 0.65$^{+0.04}_{-0.03}$&  13.488 $\pm$ 0.093 &  3.437 $\pm$ 0.064 &  35.2  \\
 25 & G168.980$+$37.738  &10.86 & 5.24 & 0.354 $\pm$ 0.053  & 2.83$^{+0.50}_{-0.37}$&   2.673 $\pm$ 0.050 & -7.221 $\pm$ 0.049 & -83.9  \\
 26 & G177.272$-$37.906  & 9.15 & 4.37 & 0.727 $\pm$ 0.073  & 1.37$^{+0.15}_{-0.12}$&  -2.731 $\pm$ 0.085 & -6.478 $\pm$ 0.062 & -40.6  \\
 27 & G179.379$+$30.743  &11.72 & 6.18 & 0.319 $\pm$ 0.106  & 3.14$^{+1.56}_{-0.78}$&  -1.407 $\pm$ 0.106 & -2.670 $\pm$ 0.081 &  -6.9  \\
 28 & G180.069$-$36.185  &12.24 & 5.91 & 0.344 $\pm$ 0.105  & 2.91$^{+1.28}_{-0.68}$&   3.782 $\pm$ 0.183 & -2.883 $\pm$ 0.111 &  60.5  \\
 29 & G180.829$+$32.784  &11.00 & 5.60 & 0.518 $\pm$ 0.095  & 1.93$^{+0.44}_{-0.30}$&  -0.383 $\pm$ 0.085 & -5.150 $\pm$ 0.068 & -24.4  \\
 30 & G181.889$-$44.366  &11.72 & 5.69 & 0.466 $\pm$ 0.089  & 2.15$^{+0.50}_{-0.34}$&   4.995 $\pm$ 0.099 & -4.558 $\pm$ 0.080 &  32.3  \\
 31 & G182.006$-$35.653  &10.67 & 5.87 & 0.856 $\pm$ 0.108  & 1.17$^{+0.17}_{-0.13}$&   6.705 $\pm$ 0.121 &  0.137 $\pm$ 0.086 &  61.0  \\
 32 & G183.614$+$31.966  & 9.22 & 5.77 & 0.665 $\pm$ 0.058  & 1.50$^{+0.14}_{-0.12}$&   3.468 $\pm$ 0.062 & -5.509 $\pm$ 0.046 &  27.5  \\
 33 & G195.025$-$53.735  &10.78 & 6.23 & 0.727 $\pm$ 0.089  & 1.38$^{+0.19}_{-0.15}$&   0.809 $\pm$ 0.088 & -3.297 $\pm$ 0.086 &  32.4  \\
 34 & G198.593$-$69.596  & 8.66 & 5.64 & 0.798 $\pm$ 0.075  & 1.25$^{+0.13}_{-0.11}$&  -1.386 $\pm$ 0.058 & -2.712 $\pm$ 0.060 &   0.4  \\
 35 & G211.919$+$50.661  & 9.13 & 4.97 & 0.830 $\pm$ 0.080  & 1.20$^{+0.13}_{-0.11}$&  -1.041 $\pm$ 0.077 &-12.503 $\pm$ 0.064 & -26.6  \\
 36 & G217.372$+$50.948  & 7.10 & 4.06 & 1.783 $\pm$ 0.050  & 0.56$^{+0.02}_{-0.02}$&  11.979 $\pm$ 0.052 &-13.905 $\pm$ 0.048 & -40.3  \\
 37 & G235.246$+$67.258  & 7.93 & 5.74 & 1.235 $\pm$ 0.086  & 0.81$^{+0.06}_{-0.05}$&   1.086 $\pm$ 0.077 & -0.127 $\pm$ 0.071 &  12.9  \\
 38 & G248.071$-$84.665  &10.32 & 5.86 & 0.460 $\pm$ 0.072  & 2.18$^{+0.40}_{-0.29}$&  12.742 $\pm$ 0.056 & -3.275 $\pm$ 0.048 & -10.8  \\
 39 & G261.694$+$46.256  &10.43 & 5.32 & 0.659 $\pm$ 0.074  & 1.52$^{+0.19}_{-0.15}$& -11.648 $\pm$ 0.086 & -4.667 $\pm$ 0.076 &  32.9  \\
 40 & G330.755$+$45.262  &11.12 & 5.71 & 0.419 $\pm$ 0.068  & 2.39$^{+0.46}_{-0.34}$&  -4.179 $\pm$ 0.075 &-11.539 $\pm$ 0.072 &  66.8  \\
 41 & G315.565$+$57.522  &10.33 & 5.60 & 0.519 $\pm$ 0.095  & 1.93$^{+0.43}_{-0.30}$&  -2.478 $\pm$ 0.127 & -0.755 $\pm$ 0.102 &  70.9  \\
 42 & G325.570$+$85.690  & 9.00 & 5.78 & 0.553 $\pm$ 0.081  & 1.81$^{+0.31}_{-0.23}$&  -9.417 $\pm$ 0.103 & -5.238 $\pm$ 0.104 &  26.1  \\
 43 & G334.109$+$36.043  &12.30 & 5.42 & 0.191 $\pm$ 0.078  & 5.23$^{+3.61}_{-1.52}$&  -2.424 $\pm$ 0.082 & -1.234 $\pm$ 0.073 & -25.4  \\
 44 & G335.504$+$35.524  &10.64 & 6.26 & 0.146 $\pm$ 0.085  & 6.83$^{+9.46}_{-2.51}$&  -4.951 $\pm$ 0.103 &  2.631 $\pm$ 0.085 & -31.2  \\
 45 & G336.532$+$38.006  &11.15 & 5.31 & 0.938 $\pm$ 0.111  & 1.07$^{+0.14}_{-0.11}$&  -4.241 $\pm$ 0.112 & -0.088 $\pm$ 0.102 &  18.1  \\
 46 & G337.373$+$32.451  &10.63 & 6.88 & 0.924 $\pm$ 0.148  & 1.08$^{+0.21}_{-0.15}$&  -2.051 $\pm$ 0.174 & -3.482 $\pm$ 0.196 &  -7.8  \\
 47 & G339.224$+$44.663  &11.69 & 5.67 & 0.204 $\pm$ 0.109  & 4.00$^{+5.62}_{-1.71}$&  -5.020 $\pm$ 0.120 &-11.514 $\pm$ 0.113 &  67.3  \\
 48 & G340.829$+$31.460  &11.15 & 5.53 & 0.419 $\pm$ 0.083  & 2.39$^{+0.59}_{-0.40}$&  -2.270 $\pm$ 0.091 & -2.023 $\pm$ 0.090 &  -3.9  \\
 49 & G345.104$+$35.879  &11.25 & 5.82 & 0.504 $\pm$ 0.076  & 1.98$^{+0.35}_{-0.26}$&  -9.532 $\pm$ 0.093 & -5.121 $\pm$ 0.068 & -59.5  \\
 50 & G349.658$+$38.897  & 9.90 & 4.19 & 0.711 $\pm$ 0.067  & 1.41$^{+0.15}_{-0.12}$&  -2.975 $\pm$ 0.076 & -3.173 $\pm$ 0.062 & -58.1  \\
 51 & G353.826$+$42.588  & 9.76 & 5.59 & 0.832 $\pm$ 0.083  & 1.20$^{+0.13}_{-0.11}$&  -9.971 $\pm$ 0.097 &-15.051 $\pm$ 0.090 &  14.9  \\
 52 & G356.642$+$59.618  &10.09 & 4.79 & 0.432 $\pm$ 0.100  & 2.32$^{+0.70}_{-0.44}$&  -0.687 $\pm$ 0.115 & -8.094 $\pm$ 0.094 &  37.2  \\
\hline                                                                                   
%\multicolumn{14}{l}{{{\bf Note:} Column 1 are ID of sources; Column 2 are Galactic coordinate notated source names; column 3 and 4 are Bayer designation names of}}\\
%\multicolumn{14}{l}{{variables and stellar types; column 5 and 6 are equatorial coordinates; column 7 and 8 are $V_{\rm LSR}$ and distance; column 9, 10, 11, 12 are}}\\
%\multicolumn{14}{l}{{peak antennas temperatures (in unit of K), line width (in unit of km s$^{-1}$), integrated flux density (in unit of K km~s$^{-1}$), and 1$\sigma$}}\\
%\multicolumn{14}{l}{{rms of spectra; column 13 are telescopes used for observations. In column 14, Y and N denote detection and non-detection of SiO maser,}}\\
%\multicolumn{14}{l}{{$\dagger$ denote sources that are still not covered by any SiO maser survey.}}\\
\end{tabular}
\end{table*}

\subsection{3D Kinematics}\label{subsec:kinematics}

In order to verify whether the elongated structure towards the
(l,b)~$\sim$~(340, 40) direction are debris of Sagittarius stellar stream or
just a coincidence of foreground disc stars due to selection effect, in this
section, we studied the kinematics of these maser-traced AGBs.

With coordinates, distances (parallaxes), proper motion and radio LSR velocity,
we can determine both the 3-dimensional location and velocity of these these
maser-traced AGB stars relative to the Sun. We followed the method of
\citet{Reid2009b} to transfer the measured heliocentric motions to a
Galactocentric reference frame. The 3D motions are described as a circular
rotation component ($\Theta_0$) plus the non-circular (peculiar) velocity
components, $U_s$, $V_s$, $W_s$, defined to be directed locally toward the
Galactic center, in the direction of rotation, and toward the north Galactic
pole.  In Table \ref{tab-3} we list the 3D velocities of these sources. Here we
adopt the A5 model of \citet{2019ApJ...885..131R}, that assuming a flat
Galactic rotation curve, with Galactic constants $R_0~=~8.15\pm0.15$~kpc,
$\Theta_0~=~236\pm7~$km~s$^{-1}$, and the solar motion values,
$U_\odot~=~10.6\pm1.2$~km~s$^{-1}$, $V_\odot~=~10.7\pm6.0~$km~s$^{-1}$,
$W_\odot~=~7.6\pm0.7$~km~s$^{-1}$.

\begin{figure*}
%\figurenum{3}
\includegraphics[width=15cm]{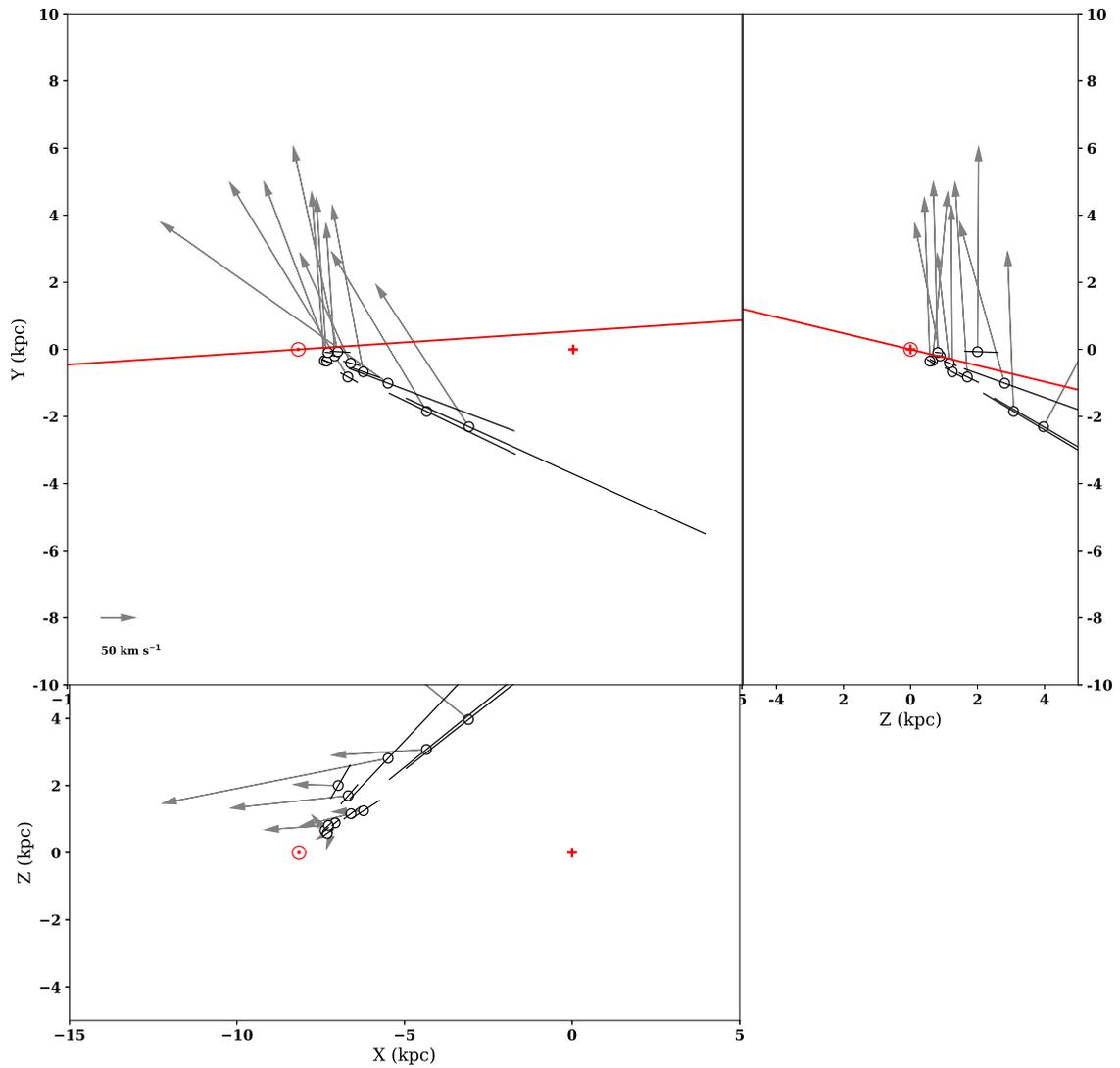}
\caption{
3D velocities of masers towards the ($l$, $b$)~$\sim$~(340, 40) direction
projected in Galactocentric Cartesian coordinates. The length of arrows denotes
velocities, with the Galactic Centre at the zero-point and the Sun at [-8.15,
0.0]~kpc. The red lines in X-Y and Y-Z plots are the intersection between the
Sgr orbital plane and X-Y and Y-Z plane. The 3D velocities suggest they are
still Galactic sources, but with a outward motion away from the Galactic center.
\label{fig-6}}
\end{figure*}

If these maser-traced AGBs are stars of stream debris, their 3D velocities
should be aligned with the Sgr orbital plane, while if they are disc stars,
their 3D velocities should be aligned with the Milky Way plane. Here, we
calculate the angles between the 3D velocities and the Sgr orbital plane
(denoted as $\hat{\theta}_{Sgr}$), and the angles between the 3D velocities and
the Galactic plane (denoted as $\hat{\theta}_{Gal}$), which are listed in the
last two columns of Table \ref{tab-3}. In Figure \ref{fig-6}, we show the
projection of the 3D velocity in Galactocentric Cartesian coordinate system. It
can be seen that the kinematics of these sources are more aligned with the
Galactic plane rather than the Sgr orbital plane. Even for the source
G339.224+44.663 with the minimum angle with respect to the Sgr orbital plane,
($\hat{\theta}_{Sgr}$~=~34$^\circ$.3~$\pm$~12$^\circ$.6), its velocity is still
more aligned with the Galactic plane
($\hat{\theta}_{Gal}$~=~9$^\circ$.2~$\pm$~7$^\circ$.5) than the Sgr orbital
plane. Therefore, kinematically, such enlongated structure traced by AGB masers
are still disk stars rather than Sgr stream debris. 

There can be two reasons that we did not detect SiO and H$_2$O masers located
in the Sgr stream. One reason can be due to sample bias, although we have
survey more than 200 sources towards sources in the Sgr orbital plane, due to
disc contaminations and sample bias, many of our targets are still thick disk
stars.  Recently, \citet{2019A+A...626A.112M} compiled a catalogue of
oxygen-rich pulsating giants in the Galactic halo and the Sgr stream, that
including more than 400 halo O-rich AGBs. The sky position indicates that
$\sim$260 stars of this catalogue with Galactic plane distance $|Z|>$~5~kpc,
are mainly members of the Sgr tidal stream. We cross matched
\citeauthor{2019A+A...626A.112M}'s catalogue with our previous SiO and H$_2$O
sample, found 19 overlaps, this means of only 5\% of
\citeauthor{2019A+A...626A.112M}'s halo and Sgr stream oxygen-rich pulsating
giants are covered by our previous survey. Another reason can be due to
limitation of sensitivity. All masers detected in our previous survey are
within 5~kpc of the Sun. Thus, for stream sources at larger distances, maser
emissions must be weaker than nearby sources, thus should be observed with
higher sensitivities.

Although the hypothesis of the Sgr stream debris is ruled out, in our previous
study on thick disk SiO masers, we succesfully identified the large-scale
peculiar motion in the Persus arm, which can be related with the Monoceros
ring, indicates the maser sources still can be potential tracers of the stream
\citep{2018MNRAS.473.3325W}.

\begin{figure*}
%\figurenum{3}
\includegraphics[width=15cm]{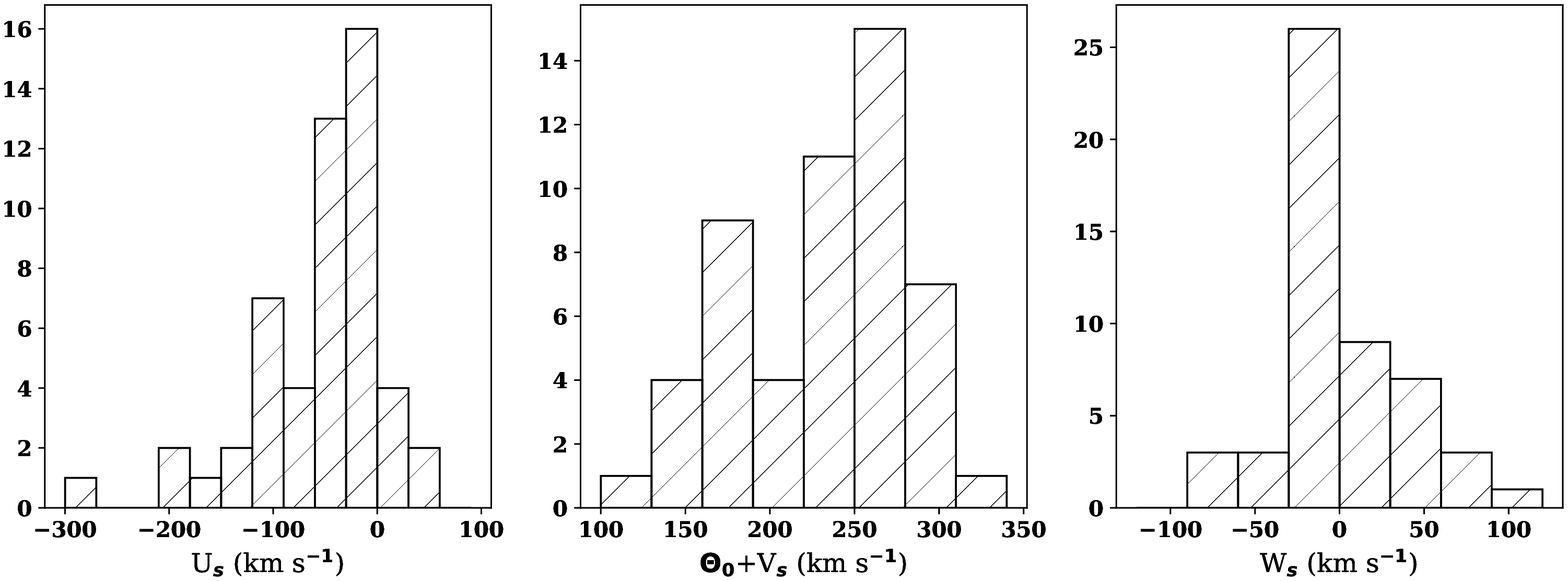}
\caption{
Histograms of velocity components $U_s$, $\Theta_0$+$V_s$, and $W_s$.
\label{fig-7}}
\end{figure*}

Further, in this study, we find a systematic motions of this maser-traced AGB
sample. In Figure \ref{fig-7}, we show the histogram of $U_s$, $\Theta_0+V_s$,
and $W_s$.  The 3D velocities of these sources indicates a remarkable outward
motions away from the Galactic center.  The $U_s$ of these sources ranges from
$-$288 to 52 km~s$^{-1}$, with mean and median values of $-$49 and $-$35~\kms. 

\citet{2000MNRAS.317..460F} studied kinematics of Miras in the Solar
neighbourhood (within 2~kpc), also found a outward motion of 75~$\pm$~18 \kms,
together with a 98~$\pm$~19~km~s$^{-1}$ lag of rotational speed.  On average,
the outward motion speed of our sample is around 30~km~s$^{-1}$, smaller than
\citet{2000MNRAS.317..460F}'s solar neighborhood Miras. While, the rotational
speed of our maser traced AGBs samples are, on average, around 100~km~s$^{-1}$
higher than that of Miras within 2~kpc of solar neighborhood. It is noted that
the galactocentric radius of our sample are generally larger than
\citet{2000MNRAS.317..460F}'s solar neighborhood Miras. Given the limited
number and the incompleteness of our maser-traced AGB sample, a further
comprehensive study on the kinematics of a complete sample of Miras and LPVs
with latest Gaia DR3 data should be necessary to yield more detailed and
confirmed conclusions.

\begin{table*}
%\footnotesize
%\setlength\tabcolsep{1pt}
%\renewcommand\arraystretch{1.23}
\caption{3D velocites of maser-traced AGBs.
\label{tab-3}}
\begin{tabular}{rcccccc}
\hline
ID &     Source & $U_s$          & $\Theta_0+V_s$  & $W_s$         & $\hat{\theta}_{Sgr}$ &  $\hat{\theta}_{Gal}$\\
   &     Name   & (km~s$^{-1}$)  & (km~s$^{-1}$)   & (km~s$^{-1}$) & ($^\circ$)         &  ($^\circ$)\\
\hline
 1  & G004.482$+$32.104 &          $-$194.00 $\pm$          75.58 & 183.12 $\pm$          30.51 & $\enspace$48.08 $\pm$          17.67 & 73.3 $\pm$ $\enspace$8.6 &          10.3 $\pm$ $\enspace$4.4 \\
 2  & G008.104$+$45.840 &          $-$129.78 $\pm$          38.01 & 241.58 $\pm$ $\enspace$5.90 & $\enspace$66.35 $\pm$          11.06 & 68.0 $\pm$ $\enspace$6.7 &          13.6 $\pm$ $\enspace$2.3 \\
 3  & G011.025$+$53.268 &       $\quad$33.28 $\pm$ $\enspace$2.53 & 257.45 $\pm$ $\enspace$3.42 &        $-$18.29 $\pm$ $\enspace$4.48 & 68.8 $\pm$ $\enspace$0.9 & $\enspace$4.0 $\pm$ $\enspace$1.0 \\
 4  & G011.159$-$41.196 &   $\enspace-$94.13 $\pm$          16.88 & 291.46 $\pm$ $\enspace$7.42 & $\enspace$21.35 $\pm$ $\enspace$4.90 & 74.3 $\pm$ $\enspace$2.4 & $\enspace$4.0 $\pm$ $\enspace$0.9 \\
 5  & G015.405$-$35.139 &       $\quad$10.54 $\pm$ $\enspace$1.56 & 277.54 $\pm$ $\enspace$4.23 & $\enspace$35.30 $\pm$ $\enspace$3.85 & 80.4 $\pm$ $\enspace$0.6 & $\enspace$7.2 $\pm$ $\enspace$0.8 \\
 6  & G019.002$-$39.495 &       $\quad-$1.21 $\pm$ $\enspace$1.99 & 144.65 $\pm$          11.10 & $\enspace-$3.49 $\pm$ $\enspace$8.63 & 73.5 $\pm$ $\enspace$3.1 & $\enspace$1.4 $\pm$ $\enspace$3.4 \\
 7  & G019.509$-$56.308 &       $\quad-$0.61 $\pm$ $\enspace$1.76 & 226.95 $\pm$ $\enspace$2.94 & $\enspace$37.64 $\pm$ $\enspace$4.39 & 82.2 $\pm$ $\enspace$0.7 & $\enspace$9.4 $\pm$ $\enspace$1.1 \\
 8  & G021.513$-$53.023 &          $-$190.27 $\pm$          30.75 & 134.83 $\pm$ $\enspace$7.15 &     $\quad$4.75 $\pm$          14.40 & 42.6 $\pm$ $\enspace$4.7 & $\enspace$1.2 $\pm$ $\enspace$3.6 \\
 9  & G022.158$+$40.858 &       $\quad-$2.07 $\pm$ $\enspace$1.97 & 163.23 $\pm$ $\enspace$5.77 & $\enspace$40.01 $\pm$ $\enspace$5.14 & 84.0 $\pm$ $\enspace$0.7 &          13.8 $\pm$ $\enspace$1.8 \\
10  & G022.943$-$31.448 &          $-$108.14 $\pm$          34.30 & 224.47 $\pm$ $\enspace$6.41 &        $-$88.84 $\pm$          27.62 & 52.2 $\pm$ $\enspace$5.3 &          19.4 $\pm$ $\enspace$5.7 \\
11  & G023.376$-$39.816 &       $\quad-$3.76 $\pm$ $\enspace$2.18 & 235.79 $\pm$ $\enspace$3.88 & $\enspace$52.41 $\pm$ $\enspace$5.67 & 83.4 $\pm$ $\enspace$0.6 &          12.5 $\pm$ $\enspace$1.3 \\
12  & G033.245$-$56.048 &   $\enspace-$49.30 $\pm$ $\enspace$3.39 & 133.74 $\pm$ $\enspace$5.63 &        $-$34.78 $\pm$ $\enspace$5.53 & 59.1 $\pm$ $\enspace$2.1 &          13.7 $\pm$ $\enspace$2.2 \\
13  & G038.070$+$66.469 &   $\enspace-$30.71 $\pm$ $\enspace$1.97 & 147.44 $\pm$ $\enspace$3.66 &        $-$10.82 $\pm$ $\enspace$4.87 & 71.1 $\pm$ $\enspace$1.8 & $\enspace$4.1 $\pm$ $\enspace$1.9 \\
14  & G041.307$-$63.037 &   $\enspace-$95.31 $\pm$ $\enspace$6.75 & 257.76 $\pm$ $\enspace$3.14 & $\enspace$38.67 $\pm$ $\enspace$4.76 & 75.0 $\pm$ $\enspace$1.3 & $\enspace$8.0 $\pm$ $\enspace$1.0 \\
15  & G041.346$-$64.747 &          $-$100.11 $\pm$          38.36 & 160.98 $\pm$          18.30 &        $-$23.69 $\pm$          17.19 & 59.5 $\pm$ $\enspace$7.2 & $\enspace$7.1 $\pm$ $\enspace$5.3 \\
16  & G045.734$-$38.770 &   $\enspace-$12.76 $\pm$ $\enspace$6.09 & 261.01 $\pm$ $\enspace$2.78 & $\enspace-$2.94 $\pm$ $\enspace$3.38 & 72.5 $\pm$ $\enspace$1.0 & $\enspace$0.6 $\pm$ $\enspace$0.7 \\
17  & G064.549$+$76.014 &          $-$163.44 $\pm$          75.76 & 114.25 $\pm$          35.32 & $\enspace$71.47 $\pm$          19.30 & 48.1 $\pm$          13.0 &          20.4 $\pm$ $\enspace$8.0 \\
18  & G085.581$-$67.859 &   $\enspace-$30.64 $\pm$ $\enspace$7.96 & 230.12 $\pm$ $\enspace$1.82 &        $-$65.38 $\pm$ $\enspace$5.70 & 60.7 $\pm$ $\enspace$1.3 &          15.7 $\pm$ $\enspace$1.3 \\
19  & G131.720$-$64.091 &       $\quad$52.72 $\pm$ $\enspace$9.03 & 267.13 $\pm$ $\enspace$5.13 &     $\quad$2.47 $\pm$ $\enspace$4.65 & 68.7 $\pm$ $\enspace$1.6 & $\enspace$0.5 $\pm$ $\enspace$1.0 \\
20  & G133.797$-$53.388 &       $\quad-$4.51 $\pm$ $\enspace$6.38 & 230.79 $\pm$ $\enspace$3.61 &        $-$12.12 $\pm$ $\enspace$5.87 & 72.0 $\pm$ $\enspace$1.5 & $\enspace$3.0 $\pm$ $\enspace$1.4 \\
21  & G141.940$-$58.536 &   $\enspace-$50.46 $\pm$ $\enspace$3.90 & 270.18 $\pm$ $\enspace$2.37 & $\enspace$29.99 $\pm$ $\enspace$4.69 & 81.1 $\pm$ $\enspace$0.9 & $\enspace$6.2 $\pm$ $\enspace$1.0 \\
22  & G149.396$-$46.550 &   $\enspace-$76.18 $\pm$          32.38 & 166.16 $\pm$          14.59 &        $-$10.21 $\pm$ $\enspace$4.50 & 68.9 $\pm$ $\enspace$4.3 & $\enspace$3.2 $\pm$ $\enspace$1.4 \\
23  & G165.616$-$40.899 &   $\enspace-$37.74 $\pm$          36.07 & 224.08 $\pm$ $\enspace$4.46 &        $-$21.80 $\pm$          10.59 & 69.0 $\pm$ $\enspace$3.4 & $\enspace$5.4 $\pm$ $\enspace$2.6 \\
24  & G166.965$-$54.751 &       $\quad-$3.07 $\pm$ $\enspace$1.60 & 185.82 $\pm$ $\enspace$3.53 & $\enspace-$9.74 $\pm$ $\enspace$4.29 & 73.2 $\pm$ $\enspace$1.3 & $\enspace$3.0 $\pm$ $\enspace$1.3 \\
25  & G168.980$+$37.738 &   $\enspace-$97.09 $\pm$          16.02 & 310.13 $\pm$ $\enspace$4.13 &        $-$12.81 $\pm$ $\enspace$6.00 & 70.6 $\pm$ $\enspace$1.7 & $\enspace$2.3 $\pm$ $\enspace$1.1 \\
26  & G177.272$-$37.906 &       $\quad-$9.87 $\pm$ $\enspace$2.35 & 292.48 $\pm$ $\enspace$4.76 & $\enspace-$8.05 $\pm$ $\enspace$4.73 & 74.8 $\pm$ $\enspace$0.9 & $\enspace$1.6 $\pm$ $\enspace$0.9 \\
27  & G179.379$+$30.743 &   $\enspace-$24.27 $\pm$          16.95 & 233.24 $\pm$ $\enspace$8.44 &        $-$17.51 $\pm$          11.93 & 71.7 $\pm$ $\enspace$2.9 & $\enspace$4.2 $\pm$ $\enspace$2.9 \\
28  & G180.069$-$36.185 &   $\enspace-$53.58 $\pm$          28.81 & 179.77 $\pm$ $\enspace$5.56 &        $-$25.14 $\pm$ $\enspace$6.06 & 64.4 $\pm$ $\enspace$3.9 & $\enspace$7.5 $\pm$ $\enspace$1.8 \\
29  & G180.829$+$32.784 &   $\enspace-$35.09 $\pm$          10.17 & 258.25 $\pm$ $\enspace$4.43 &        $-$12.61 $\pm$ $\enspace$3.60 & 73.1 $\pm$ $\enspace$0.9 & $\enspace$2.8 $\pm$ $\enspace$0.8 \\
30  & G181.889$-$44.366 &   $\enspace-$58.63 $\pm$          16.04 & 211.21 $\pm$ $\enspace$3.87 &        $-$17.90 $\pm$ $\enspace$3.94 & 68.1 $\pm$ $\enspace$2.3 & $\enspace$4.6 $\pm$ $\enspace$1.0 \\
31  & G182.006$-$35.653 &   $\enspace-$14.96 $\pm$ $\enspace$3.43 & 171.00 $\pm$ $\enspace$4.71 &        $-$12.14 $\pm$ $\enspace$4.49 & 72.4 $\pm$ $\enspace$1.5 & $\enspace$4.0 $\pm$ $\enspace$1.5 \\
32  & G183.614$+$31.966 &   $\enspace-$33.76 $\pm$ $\enspace$4.00 & 231.90 $\pm$ $\enspace$4.54 & $\enspace$39.02 $\pm$ $\enspace$3.14 & 83.6 $\pm$ $\enspace$0.9 & $\enspace$9.5 $\pm$ $\enspace$0.8 \\
33  & G195.025$-$53.735 &   $\enspace-$15.63 $\pm$ $\enspace$2.88 & 235.37 $\pm$ $\enspace$3.62 &        $-$33.85 $\pm$ $\enspace$4.16 & 68.3 $\pm$ $\enspace$1.0 & $\enspace$8.2 $\pm$ $\enspace$1.0 \\
34  & G198.593$-$69.596 &$\quad\enspace$2.02 $\pm$ $\enspace$1.40 & 262.29 $\pm$ $\enspace$2.75 & $\enspace-$9.19 $\pm$ $\enspace$4.86 & 74.1 $\pm$ $\enspace$1.0 & $\enspace$2.0 $\pm$ $\enspace$1.1 \\
35  & G211.919$+$50.661 &   $\enspace-$51.35 $\pm$ $\enspace$7.25 & 284.37 $\pm$ $\enspace$3.72 &        $-$26.41 $\pm$ $\enspace$4.41 & 69.2 $\pm$ $\enspace$1.0 & $\enspace$5.2 $\pm$ $\enspace$0.9 \\
36  & G217.372$+$50.948 &       $\quad-$6.30 $\pm$ $\enspace$2.31 & 308.19 $\pm$ $\enspace$2.98 & $\enspace-$9.44 $\pm$ $\enspace$4.01 & 74.7 $\pm$ $\enspace$0.8 & $\enspace$1.8 $\pm$ $\enspace$0.8 \\
37  & G235.246$+$67.258 &$\quad\enspace$7.21 $\pm$ $\enspace$1.86 & 255.12 $\pm$ $\enspace$1.74 & $\enspace$20.78 $\pm$ $\enspace$4.68 & 80.4 $\pm$ $\enspace$1.0 & $\enspace$4.6 $\pm$ $\enspace$1.1 \\
38  & G248.071$-$84.665 &   $\enspace-$90.47 $\pm$          17.24 & 162.60 $\pm$          15.20 & $\enspace$20.98 $\pm$ $\enspace$5.44 & 62.7 $\pm$ $\enspace$4.9 & $\enspace$6.4 $\pm$ $\enspace$1.7 \\
39  & G261.694$+$46.256 &   $\enspace-$59.09 $\pm$ $\enspace$5.59 & 209.31 $\pm$ $\enspace$4.97 &        $-$17.36 $\pm$ $\enspace$7.36 & 63.8 $\pm$ $\enspace$1.7 & $\enspace$4.5 $\pm$ $\enspace$1.9 \\
40  & G330.755$+$45.262 &          $-$138.04 $\pm$          24.75 & 308.64 $\pm$ $\enspace$3.62 &        $-$18.49 $\pm$          13.43 & 58.2 $\pm$ $\enspace$3.4 & $\enspace$3.1 $\pm$ $\enspace$2.3 \\
41  & G315.565$+$57.522 &   $\enspace-$35.38 $\pm$ $\enspace$5.26 & 276.62 $\pm$ $\enspace$2.48 & $\enspace$62.18 $\pm$ $\enspace$4.37 & 81.1 $\pm$ $\enspace$1.0 &          12.6 $\pm$ $\enspace$0.9 \\
42  & G325.570$+$85.690 &   $\enspace-$71.94 $\pm$          13.08 & 207.06 $\pm$ $\enspace$6.80 & $\enspace$25.07 $\pm$ $\enspace$5.07 & 72.8 $\pm$ $\enspace$3.0 & $\enspace$6.5 $\pm$ $\enspace$1.3 \\
43  & G334.109$+$36.043 &   $\enspace-$36.66 $\pm$          34.72 & 273.42 $\pm$          33.20 & $\enspace-$8.91 $\pm$ $\enspace$3.66 & 58.9 $\pm$ $\enspace$6.0 & $\enspace$1.9 $\pm$ $\enspace$0.8 \\
44  & G335.504$+$35.524 &       $\quad$16.21 $\pm$          12.78 & 251.46 $\pm$          62.82 &          112.02 $\pm$          42.38 & 59.0 $\pm$ $\enspace$4.3 &          24.6 $\pm$          10.3 \\
45  & G336.532$+$38.006 &       $\quad-$7.54 $\pm$ $\enspace$2.76 & 251.27 $\pm$ $\enspace$3.73 & $\enspace$22.63 $\pm$ $\enspace$3.44 & 81.6 $\pm$ $\enspace$0.8 & $\enspace$5.1 $\pm$ $\enspace$0.8 \\
46  & G337.373$+$32.451 &       $\quad-$3.56 $\pm$ $\enspace$4.21 & 243.58 $\pm$ $\enspace$4.29 & $\enspace-$8.28 $\pm$ $\enspace$3.34 & 74.5 $\pm$ $\enspace$0.8 & $\enspace$1.9 $\pm$ $\enspace$0.8 \\
47  & G339.224$+$44.663 &          $-$287.49 $\pm$          99.77 & 296.41 $\pm$          95.83 &        $-$67.09 $\pm$          51.34 & 34.3 $\pm$          12.6 & $\enspace$9.2 $\pm$ $\enspace$7.5 \\
48  & G340.829$+$31.460 &   $\enspace-$19.21 $\pm$ $\enspace$9.18 & 250.70 $\pm$ $\enspace$6.61 & $\enspace-$1.60 $\pm$ $\enspace$3.03 & 74.5 $\pm$ $\enspace$1.1 & $\enspace$0.4 $\pm$ $\enspace$0.7 \\
49  & G345.104$+$35.879 &   $\enspace-$64.49 $\pm$          14.41 & 167.41 $\pm$ $\enspace$6.37 &        $-$18.35 $\pm$ $\enspace$3.97 & 61.5 $\pm$ $\enspace$3.1 & $\enspace$5.9 $\pm$ $\enspace$1.3 \\
40  & G349.658$+$38.897 &       $\quad-$7.24 $\pm$ $\enspace$3.02 & 196.91 $\pm$ $\enspace$3.98 &        $-$37.71 $\pm$ $\enspace$3.32 & 65.6 $\pm$ $\enspace$1.0 &          10.8 $\pm$ $\enspace$1.0 \\
51  & G353.826$+$42.588 &   $\enspace-$91.27 $\pm$          10.88 & 255.13 $\pm$ $\enspace$3.86 & $\enspace-$6.52 $\pm$ $\enspace$3.94 & 67.8 $\pm$ $\enspace$1.7 & $\enspace$1.4 $\pm$ $\enspace$0.8 \\
52  & G356.642$+$59.618 &   $\enspace-$62.88 $\pm$          23.29 & 305.39 $\pm$          14.34 &     $\quad$1.58 $\pm$ $\enspace$9.87 & 73.9 $\pm$ $\enspace$2.6 & $\enspace$0.3 $\pm$ $\enspace$1.8 \\
\hline  
%\multicolumn{14}{l}{{{\bf Note:} Column 1 are ID of sources; Column 2 are Galactic coordinate notated source names; column 3 and 4 are Bayer designation names of}}\\
%\multicolumn{14}{l}{{variables and stellar types; column 5 and 6 are equatorial coordinates; column 7 and 8 are $V_{\rm LSR}$ and distance; column 9, 10, 11, 12 are}}\\
%\multicolumn{14}{l}{{peak antennas temperatures (in unit of K), line width (in unit of km s$^{-1}$), integrated flux density (in unit of K km~s$^{-1}$), and 1$\sigma$}}\\
%\multicolumn{14}{l}{{rms of spectra; column 13 are telescopes used for observations. In column 14, Y and N denote detection and non-detection of SiO maser,}}\\
%\multicolumn{14}{l}{{$\dagger$ denote sources that are still not covered by any SiO maser survey.}}\\
\end{tabular}
\end{table*}

\section{Summary} \label{sec:summary}
We conducted an water maser survey towards 176 O-rich AGBs towards the Sgr
orbital plane. In total, we detected maser emissions in 21 AGBs, of which 20
were new detection. Together with previous SiO maser survey data and Gaia EDR3
data, we studied the 3D location and kinematics of this maser-traced AGB
sample. We found an elongated sources distributions towards the ($l$,
$b$)~$\sim$~(340$^{\circ}$, 40$^{\circ}$) direction. However, the 3D kinematics
of this structure is more aligned with the Galactic plane rather than the Sgr
orbital plane. Kinematically, such maser-traced AGB stars has a systematic
outward motions (30~\kms) away from the Galactic center. However, we found no
systematic lags of rotational speed which was reported in
\citet{2000MNRAS.317..460F}'s study on Miras of solar neighborhood.

\section*{Acknowledgements}
% Entry for the table of contents, for this guide only
\addcontentsline{toc}{section}{Acknowledgements}
We would like to thank Shinji Horiuchi from CSIRO for his kindly support on
the observations and data reduction of Tidbinbilla 70m telescope. We would like
to thank NRO45m staffs for their helps and supports during observations. The
Nobeyama 45-m radio telescope is operated by Nobeyama Radio Observatory, a
branch of National Astronomical Observatory of Japan. The 70-m radio telescope
are operated by the Canberra Deep Space Communication Complex, part of NASA's
Deep Space Network. This work has made use of data from the European Space Agency (ESA) mission
{\it Gaia} (\url{https://www.cosmos.esa.int/gaia}), processed by the {\it Gaia}
Data Processing and Analysis Consortium (DPAC,
\url{https://www.cosmos.esa.int/web/gaia/dpac/consortium}). Funding for the DPAC
has been provided by national institutions, in particular the institutions
participating in the {\it Gaia} Multilateral Agreement.

\section*{Data availability}
The original maser survey data underlying this article are available in CAS Cloud Box, at 
{\it{https://pan.cstcloud.cn/s/bLYIDwtTZE}}, with a password {\it{Sagi}}.

\newpage
%\bibliographystyle{apj}
%\bibliography{ref_ywwu}

\appendix
%\appendix 
\setcounter{table}{0}
\setcounter{figure}{0}
\renewcommand{\thetable}{A\arabic{table}}
\renewcommand{\thefigure}{A\arabic{figure}}
\newpage
\begin{table*}
\caption{List of observed Sources}
\label{tab-A1}
\begin{tabular}{ccccrrrcc}
\hline
Source & WISE      &  Other  & Star   &  Period & Maser      \\
Name   & Name      &  Name   & Type   &  (days) & Detection \\
\hline
G011.159-41.196 & J210436.73-331647.9 &  X Mic                  &Mi*     &  239.8  &  Y \\
G012.230-32.997 & J202729.16-304837.0 &  V5556 Sgr              &Mi*     &  233.0  &  N \\
G012.901-63.916 & J225330.68-325539.9 &  SS PsA                 &Mi*     &  195.0  &  N \\
G015.405-35.139 & J204002.99-284730.1 &  R Mic                  &Mi*     &  138.6  &  N \\
G016.830-45.308 & J212643.96-295105.5 &  S Mic                  &Mi*     &  209.6  &  N \\
G019.002-39.495 & J210220.75-270514.6 &  RR Cap                 &Mi*     &  277.5  &  N \\
G019.509-56.308 & J221800.18-293613.5 &  R PsA                  &Mi*     &  292.3  &  N \\
G021.290-47.410 & J213841.86-271234.0 &  RV PsA                 &Mi*     &  361.0  &  N \\
G021.513-53.023 & J220345.83-280303.3 &  S PsA                  &Mi*     &  271.0  &  Y \\
G022.943-31.448 & J203234.16-214126.5 &  RU Cap                 &Mi*     &  347.3  &  N \\
G023.376-39.816 & J210736.63-235513.4 &  V Cap                  &Mi*     &  275.7  &  Y \\
G026.645-39.261 & J210833.06-212051.5 &  X Cap                  &Mi*     &  218.9  &  N \\
G028.805-31.577 & J204032.07-170328.1 &  TX Cap                 &Mi*     &  199.0  &  N \\
G032.106-32.402 & J204808.58-144701.0 &  U Cap                  &Mi*     &  203.0  &  N \\
G033.078-37.937 & J211037.51-161024.8 &  Z Cap                  &Mi*     &  181.4  &  N \\
G033.245-56.048 & J222312.94-220325.5 &  RT Aqr                 &Mi*     &  247.0  &  Y \\
G033.519-53.484 & J221250.92-210951.8 &  AQ Aqr                 &Mi*     &  237.1  &  N \\
G035.634-40.082 & J212200.72-150932.4 &  T Cap                  &Mi*     &  271.9  &  N \\
G037.206-55.581 & J222413.45-194742.7 &  AV Aqr                 &Mi*     &  250.7  &  N \\
G038.666-42.354 & J213422.91-135829.2 &  Y Cap                  &Mi*     &  395.0  &  N \\
G041.307-63.037 & J225706.46-202035.7 &  S Aqr                  &Mi*     &  279.3  &  Y \\
G041.346-64.747 & J230400.56-205424.0 &  MN Aqr                 &Mi*     &  285.0  &  Y \\
G045.006-52.540 & J221954.43-142406.9 &  SS Aqr                 &Mi*     &  202.2  &  N \\
G045.032-36.798 & J212303.66-070629.4 &  RZ Aqr                 &Mi*     &  396.0  &  N \\
G045.734-38.770 & J213106.49-073420.4 &  HY Aqr                 &Mi*     &  311.7  &  N \\
G055.342-64.356 & J231324.09-151916.0 &  UX Aqr                 &Mi*     &  320.0  &  N \\
G077.780-73.063 & J000207.53-144033.9 &  W Cet                  &Mi*     &  352.0  &  N \\
G085.581-67.859 & J235754.06-085731.2 &  V Cet                  &Mi*     &  260.0  &  N \\
G131.720-64.091 & J010645.20-012851.8 &  Z Cet                  &Mi*     &  184.0  &  N \\
G133.797-53.388 & J011734.56+085551.9 &  S Psc                  &Mi*     &  404.6  &  N \\
G134.760-49.278 & J012258.47+125204.0 &  U Psc                  &Mi*     &  172.8  &  N \\
G141.940-58.536 & J013038.34+025252.4 &                         &        &         &  N \\
G149.396-46.550 & J020437.67+123136.9 &  S Ari                  &Mi*     &  291.0  &  N \\
G165.616-40.899 & J025727.51+111805.2 &  YZ Ari                 &Mi*     &  447.0  &  N \\
G168.980+37.738 & J084049.54+500812.0 &  X UMa                  &Mi*     &  249.0  &  Y \\
G179.379+30.743 & J080503.69+405908.1 &                         &        &         &  Y \\
G180.829+32.784 & J081646.88+400753.2 &  W Lyn                  &Mi*     &  295.2  &  Y \\
G183.613+31.966 & J081450.57+374015.4 &  RT Lyn                 &Mi*     &  394.6  &  N \\
G195.025-53.735 & J031153.13-115232.3 &  SS Eri                 &Mi*     &  319.0  &  N \\
G198.593-69.596 & J021600.03-203110.7 &  RY Cet                 &Mi*     &  364.0  &  Y \\
G206.317+31.191 & J083546.31+185344.7 &  U Cnc                  &Mi*     &  305.0  &  N \\
G211.919+50.661 & J100001.98+211543.9 &  V Leo                  &Mi*     &  273.3  &  Y \\
G212.108+46.532 & J094325.67+195139.9 &  RS Leo                 &Mi*     &  208.2  &  N \\
G248.071-84.665 & J011136.37-300629.4 &  U Scl                  &Mi*     &  333.7  &  Y \\
G315.565+57.522 & J131830.41-044105.0 &  VY Vir                 &Mi*     &  277.0  &  Y \\
\hline
\multicolumn{6}{l}{Column 1 are Galactic coordinate notated source names; column 2 are equatorial}\\
\multicolumn{6}{l}{coordinates notated WISE name; column 3 are Bayerdesignation names of variables}\\
\multicolumn{6}{l}{column 4 are stellar types; column 5 are periods; column 6 denote detections of}\\
\multicolumn{6}{l}{maser survey.}

\end{tabular}
\end{table*}

\begin{table*}
\contcaption{List of observed Sources}
\begin{tabular}{ccccrrrcc}
\hline
Source & WISE      &  Other  & Star   &  Period & Maser      \\
Name   & Name      &  Name   & Type   &  (days) & Detection \\
\hline
G120.894-64.174 & J004753.14-011858.7 &  SX Cet                 &sr      &  200.0  &  N \\
G107.659-63.974 & J002452.54-015335.4 &  PB 5937                &cv*     &         &  N \\
G107.359-64.023 & J002424.58-015818.4 &  DY Psc                 &by*     &         &  N \\
G071.783-70.795 & J235007.19-141750.2 &                         &Mi*?    &   84.7  &  N \\
G069.300-72.233 & J235214.54-155117.2 &  Z Aqr                  &SRA     &  136.6  &  N \\
G073.394-77.357 & J001057.96-183423.3 &  AC Cet                 &LB      &         &  N \\
G083.632-79.140 & J002230.88-183245.1 &                         &Mi*     &  197.0  &  N \\
G094.565-80.456 & J003221.61-183908.4 &  ET Cet                 &LP?     &   64.1  &  N \\
G046.841-80.767 & J001159.74-243359.4 &  GL Cet                 &SRB     &  101.0  &  N \\
G025.175-76.128 & J234855.61-280750.1 &                         &        &         &  N \\
G003.139-73.955 & J234405.91-340511.3 &                         &Mi*?    &   61.6  &  N \\
G356.903-74.722 & J235053.76-351748.6 &                         &Mi*?    &  159.0  &  N \\
G346.589-78.389 & J001217.03-351113.5 &  CO Scl                 &LB      &   64.4  &  N \\
G181.209-73.392 & J015033.86-173900.9 &  DH Cet                 &SRB     &  212.0  &  N \\
G184.630-72.876 & J015445.94-180905.4 &                         &Mi*?    &  337.0  &  N \\
G158.558-71.111 & J013547.91-112230.1 &  FY Cet                 &SRB     &  211.8  &  N \\
G155.834-67.648 & J013935.49-075421.8 &                         &Mi*?    &   98.0  &  N \\
G144.537-70.053 & J012037.11-082452.5 &  CU Cet                 &SRB     &  151.0  &  N \\
G147.548-60.663 & J013832.50-000343.6 &                         &SRA     &  169.0  &  N \\
G146.496-59.340 & J013830.11+012140.0 &  SW Cet                 &LB      &   53.9  &  N \\
G135.687-52.241 & J012259.10+095050.1 &                         &Mi*?    &  115.7  &  N \\
G150.794-47.560 & J020627.29+111246.1 &  CS Ari                 &SRB     &   67.6  &  N \\
G156.088-52.393 & J020946.82+052141.7 &                         &Mi*?    &   78.7  &  N \\
G159.820-47.007 & J022917.65+084408.4 &                         &Mi*?    &   81.5  &  N \\
G163.265-46.216 & J023900.42+080341.2 &                         &Mi*?    &   72.0  &  N \\
G165.494-43.684 & J025014.09+090916.2 &                         &Mi*?    &   35.1  &  N \\
G158.741-40.144 & J024144.10+145612.3 &                         &Mi*?    &   72.5  &  N \\
G159.484-38.173 & J024812.38+161628.2 &  BD Ari                 &SRB     &  192.0  &  N \\
G156.914-37.397 & J024246.87+180113.3 &                         &Mi*?    &   88.5  &  N \\
G162.544-33.901 & J030658.77+182044.1 &                         &Mi*?    &   65.6  &  N \\
G169.816-33.691 & J032659.91+143957.0 &                         &SR      &  376.0  &  N \\
G179.553-33.716 & J035006.63+085209.0 &                         &Mi*?    &  282.0  &  N \\
G180.069-36.185 & J034343.89+065530.5 &  V1083 Tau              &Mi*     &  343.0  &  Y \\
G182.006-35.653 & J034927.68+060440.3 &  V1191 Tau              &Mi*     &  338.0  &  N \\
G177.272-37.906 & J033232.89+072532.1 &                         &Mi*?    &  158.6  &  Y \\
G173.516-38.103 & J032335.70+092354.9 &                         &Mi*?    &   74.9  &  N \\
G168.016-39.975 & J030537.06+105217.7 &                         &Mi*?    &   70.2  &  N \\
G166.990-37.284 & J031007.92+132712.8 &  ST Ari                 &SRB     &   96.0  &  N \\
G170.281-45.769 & J025530.81+052315.1 &                         &M:      &  185.0  &  N \\
G170.765-44.664 & J025928.69+055931.6 &                         &Mi*?    &  201.1  &  N \\
G181.889-44.367 & J032231.60+003147.9 &                         &Mi*     &  213.3  &  Y \\
G182.059-45.211 & J032015.56-000628.9 &                         &Mi*?    &  335.0  &  N \\
G180.468-50.695 & J030050.91-025244.5 &  CV Eri                 &LB:     &         &  N \\
G177.437-56.493 & J023847.79-052650.6 &  KL Cet                 &SR      &   81.0  &  N \\
G191.808-60.569 & J024407.33-135131.4 &                         &        &         &  N \\
\hline
\multicolumn{6}{l}{Column 1 are Galactic coordinate notated source names; column 2 are equatorial}\\
\multicolumn{6}{l}{coordinates notated WISE name; column 3 are Bayerdesignation names of variables}\\
\multicolumn{6}{l}{column 4 are stellar types; column 5 are periods; column 6 denote detections of}\\
\multicolumn{6}{l}{maser survey.}
\end{tabular}
\end{table*}

\begin{table*}
\contcaption{List of observed Sources}
\begin{tabular}{ccccrrrcc}
\hline
Source & WISE      &  Other  & Star   &  Period & Maser      \\
Name   & Name      &  Name   & Type   &  (days) & Detection \\
\hline
G203.330+30.788 & J083022.53+210927.3 &                         &Mi*?    &   49.8  &  N \\
G204.101+32.634 & J083846.63+210932.6 &  UV Cnc                 &SRB     &  148.5  &  N \\
G204.907+32.225 & J083806.75+202250.2 &  DK Cnc                 &SRB     &  100.0  &  N \\
G210.261+35.437 & J085725.83+172051.9 &                         &Mi*     &  391.0  &  N \\
G211.285+38.404 & J091016.62+173922.4 &                         &Mi*?    &   92.4  &  N \\
G209.815+40.043 & J091505.22+191737.8 &                         &SR      &  261.2  &  N \\
G209.312+39.791 & J091331.88+193422.8 &                         &SRA     &  261.5  &  N \\
G208.315+41.183 & J091805.45+204432.6 &                         &L:      &  130.0  &  N \\
G214.232+43.058 & J093151.13+171505.4 &                         &Mi*     &  179.0  &  N \\
G221.703+45.061 & J094816.60+130653.8 &                         &        &         &  N \\
G223.695+47.335 & J095908.69+125155.4 &                         &Mi*?    &   77.5  &  N \\
G223.754+51.667 & J101516.72+144259.4 &                         &Mi*?    &   78.1  &  N \\
G218.786+51.178 & J100814.78+172130.5 &  DD Leo                 &LPV     &  110.9  &  N \\
G217.372+50.948 & J100558.79+180604.9 &                         &LPV?    &  271.0  &  Y \\
G238.855+56.995 & J105211.03+094855.2 &                         &Mi*?    &  490.0  &  N \\
G329.032+52.959 & J135320.96-064447.5 &  AI Vir                 &Mi*?    &R  80.5  &  N \\
G333.090+53.064 & J140204.43-053715.9 &  AB Vir                 &SR      &  313.0  &  N \\
G337.755+51.213 & J141544.46-055206.3 &  CF Vir                 &Mi*     &  226.5  &  N \\
G331.573+49.548 & J140453.43-091141.2 &  RR Vir                 &Mi*     &  217.9  &  N \\
G335.646+44.462 & J142429.53-122507.3 &                         &        &         &  N \\
G339.224+44.663 & J143259.86-105603.2 &  KS Lib                 &Mi*     &  371.0  &  Y \\
G339.938+43.686 & J143654.68-112840.8 &                         &Mi*?    &R 101.5  &  N \\
G344.310+38.251 & J150049.25-140045.6 &                         &Mi*?    &R 106.7  &  N \\
G349.658+38.897 & J151223.62-105151.6 &                         &Mi*?    &   63.6  &  Y \\
G347.902+34.970 & J151835.76-144503.3 &                         &Mi*?    &  230.3  &  N \\
G348.565+34.628 & J152113.37-143912.9 &                         &Mi*?    &R 131.4  &  N \\
G351.579+32.192 & J153541.94-144516.2 &                         &Mi*?    &   56.1  &  N \\
G350.237+30.144 & J153805.24-170154.1 &  EK Lib                 &SR      &  179.9  &  N \\
G345.966+35.014 & J151325.77-154359.6 &                         &Mi*     &  270.0  &  N \\
G345.104+35.879 & J150854.49-152951.0 &  TT Lib                 &Mi*     &  283.0  &  Y \\
G342.197+32.029 & J151044.35-200108.3 &  T Lib                  &Mi*     &  237.5  &  N \\
G340.829+31.460 & J150810.65-211000.2 &  YY Lib                 &Mi*     &  230.2  &  N \\
G341.040+30.544 & J151107.51-214802.9 &  AD Lib                 &SRB     &  292.1  &  N \\
G338.587+32.039 & J150001.99-214704.0 &                         &Mi*     &  353.0  &  N \\
G338.451+33.182 & J145653.67-205353.1 &                         &Mi*?    &  528.6  &  N \\
G336.386+35.874 & J144436.97-193228.2 &  TW Lib                 &Mi*     &  214.8  &  N \\
G334.109+36.043 & J143729.13-201941.2 &  LY Lib                 &Mi*     &  284.0  &  N \\
G333.047+39.864 & J142638.13-172235.7 &                         &Mi*?    &R  82.0  &  N \\
G338.952+42.021 & J143807.15-131608.8 &                         &SR      &  136.0  &  N \\
G326.227+43.955 & J140016.12-154652.6 &                         &Mi*?    &  114.2  &  N \\
G005.320-60.346 & J223827.91-365342.3 &  CY Gru                 &LB      &   83.3  &  N \\
G005.009-62.225 & J224750.07-364208.6 &                         &Mi*?    &  169.8  &  N \\
G006.641-64.182 & J225638.18-353606.3 &  SU PsA                 &Mi*     &  224.0  &  N \\
G026.741-68.997 & J231647.58-272333.8 &                         &        &         &  N \\
\hline
\multicolumn{6}{l}{Column 1 are Galactic coordinate notated source names; column 2 are equatorial}\\
\multicolumn{6}{l}{coordinates notated WISE name; column 3 are Bayerdesignation names of variables}\\
\multicolumn{6}{l}{column 4 are stellar types; column 5 are periods; column 6 denote detections of}\\
\multicolumn{6}{l}{maser survey.}
\end{tabular}
\end{table*}

\begin{table*}
\contcaption{List of observed Sources}
\begin{tabular}{ccccrrrcc}
\hline
Source & WISE      &  Other  & Star   &  Period & Maser      \\
Name   & Name      &  Name   & Type   &  (days) & Detection \\
\hline
G028.838-67.773 & J231141.72-262945.4 &                         &Mi*?    &   77.8  &  N \\
G030.300-65.497 & J230203.56-253511.0 &                         &Mi*?    &   50.5  &  N \\
G026.656-64.777 & J225753.94-265954.4 &  XY PsA                 &SR:     &  202.7  &  N \\
G030.486-59.109 & J223429.69-241518.1 &                         &Mi*?    &   69.1  &  N \\
G026.546-53.173 & J220653.66-250628.1 &                         &Mi*     &  328.0  &  N \\
G025.487-60.974 & J224039.35-270237.0 &                         &        &         &  N \\
G015.996-59.189 & J223054.39-313744.2 &                         &Mi*?    &  133.8  &  N \\
G014.046-39.222 & J205715.94-304551.8 &  CT Mic                 &LPV     &         &  N \\
G009.843-45.833 & J212618.51-344606.4 &                         &Mi*?    &   75.9  &  N \\
G009.444-47.911 & J213620.28-351047.2 &                         &        &         &  N \\
G013.301-48.406 & J213937.40-323900.4 &                         &Mi*?    &  181.0  &  N \\
G014.066-48.985 & J214234.98-321210.6 &  SY PsA                 &Mi*     &  332.0  &  N \\
G019.000-48.457 & J214206.71-285426.8 &                         &LPV     &  137.4  &  N \\
G023.134-46.163 & J213427.90-254232.6 &  XX PsA                 &LPV?    &  184.5  &  N \\
G028.021-47.167 & J214225.82-224303.7 &                         &Mi*?    &  105.3  &  N \\
G031.942-44.991 & J213704.84-192757.5 &                         &        &         &  N \\
G025.952-42.117 & J211937.45-224226.9 &  CH Cap                 &SRA     &  180.0  &  N \\
G018.685-42.378 & J211445.48-275841.8 &  CV Mic                 &Mi*     &  292.0  &  N \\
G023.139-38.703 & J210242.82-234654.8 &  CE Cap                 &LPV?    &   75.7  &  N \\
G026.343-35.418 & J205239.24-202000.2 &  BX Cap                 &SRA     &  158.0  &  N \\
G025.457-33.119 & J204222.26-201500.3 &                         &Mi*?    &  306.0  &  N \\
G023.291-34.188 & J204409.74-221812.9 &  CC Cap                 &SRB     &  158.0  &  N \\
G022.620-33.670 & J204116.20-224004.0 &                         &Mi*?    &  171.2  &  N \\
G019.225-35.150 & J204348.80-254656.2 &                         &Mi*?    &  216.5  &  N \\
G016.099-35.257 & J204113.07-281621.7 &  CQ Mic                 &LPV?    &  142.4  &  N \\
G013.595-34.912 & J203722.90-301025.9 &  RT Mic                 &Mi*     &  113.6  &  N \\
G013.274-32.111 & J202432.95-294402.6 &  V6592 Sgr              &Mi*     &  215.3  &  N \\
G014.702-33.029 & J203003.35-284849.6 &                         &SRB     &  173.0  &  N \\
G014.973-32.742 & J202904.35-283106.2 &  RS Mic                 &Mi*     &  224.0  &  N \\
G015.054-30.568 & J201940.82-275055.3 &                         &Mi*?    &   74.5  &  N \\
G017.724-30.348 & J202147.45-253507.2 &  EP Cap                 &SRB     &  253.0  &  N \\
G020.708-31.502 & J203006.77-233041.5 &  AY Cap                 &SRB     &  194.0  &  N \\
G021.343-32.293 & J203407.64-231458.5 &  AK Cap                 &SRB     &   92.0  &  N \\
G021.776-30.351 & J202640.44-221617.9 &                         &Mi*?    &  132.4  &  N \\
G039.040-42.707 & J213610.46-135204.9 &  UU Cap                 &SRB     &  100.0  &  N \\
G038.444-46.662 & J215027.92-155028.3 &  AA Cap                 &SRB     &   73.2  &  N \\
G039.598-46.255 & J215013.47-145701.0 &                         &Mi*?    &   53.0  &  N \\
G041.287-48.941 & J220218.18-150014.8 &                         &Mi*?    &  108.2  &  N \\
G041.058-50.181 & J220644.98-153840.2 &  BM Aqr                 &SRB     &   55.6  &  N \\
G041.037-51.361 & J221113.35-160747.1 &  YY Aqr                 &SRB     &  197.0  &  N \\
G036.242-55.590 & J222330.38-201849.2 &  KU Aqr                 &LB      &         &  N \\
G037.206-55.581 & J222413.45-194742.7 &  AV Aqr                 &Mi*     &  250.7  &  N \\
\hline
\multicolumn{6}{l}{Column 1 are Galactic coordinate notated source names; column 2 are equatorial}\\
\multicolumn{6}{l}{coordinates notated WISE name; column 3 are Bayerdesignation names of variables}\\
\multicolumn{6}{l}{column 4 are stellar types; column 5 are periods; column 6 denote detections of}\\
\multicolumn{6}{l}{maser survey.}
\end{tabular}
\end{table*}
% Don't change these lines
\bsp	% typesetting comment
\label{lastpage}
\end{document}